\documentstyle[12pt,aaspp4]{article}

\def\O{\Omega}
\def\teff{T_{eff}}
\def\cm{{\rm\ cm}} 
\def\xec{\times 10^{11} {\rm\ cm}}
\def\xtc{\times 10^{10} \cm}
\def\pers{~{\rm s}^{-1}} %requires $...$ in text
\def\xfs{\times 10^{-5} \pers}
\def\xss{\times 10^{-6} \pers}
\def\ergs{{\rm ~ergs~s^{-1}}}

\def\msun{{\,\rm M_{\odot}}}
\def\rsun{{\,\rm R_{\odot}}}
\def\lsun{{\,\rm L_{\odot}}}
\def\msyr{{\msun}~{\rm yr}^{-1}}
\def\O{\Omega}

\def\ok{\Omega_K}
\def\oms{\Omega_*} %requires $...$ in text
\def\omseq{\Omega_{*,eq}}
\def\omseqf{\Omega_{*,eq,FU}}
\def\omseqt{\Omega_{*,eq,TT}}

\def\omko{\Omega_K(R_{out})}
\def\omks{\Omega_K(R_{_*})}
\def\omk2i{\Omega_K^2(R_{in})}
\def\omk2o{\Omega_K^2(R_{out})}
\def\omk2s{\Omega_K^2(R_{_*})}
\def\md{\dot M}
\def\ms{M_{_*}}
\def\rs{R_{_*}}

\def\ro{R_{out}}
\def\hs{H_{_*}}
\def\vs{v_{R,*}}

\def\et{{\it et al.}\ }
\def\D{\Delta}
\def\dmf{\D M_{FU}}
\def\dmt{\D M_{TT}}

\def\sles{\lower2pt\hbox{$\buildrel {\scriptstyle <}
   \over {\scriptstyle\sim}$}}
\def\sgreat{\lower2pt\hbox{$\buildrel {\scriptstyle >}
   \over {\scriptstyle\sim}$}}
\begin{document}

\title{CAN FU ORIONIS OUTBURSTS REGULATE THE ROTATION RATES OF T TAURI
STARS?} 
\author{Robert Popham}
\affil{Harvard-Smithsonian Center for Astrophysics}
\authoraddr{MS 51, 60 Garden St., Cambridge, MA 02138}

\begin{abstract}

We propose that FU Orionis outbursts may play an important role in
maintaining the slow rotation of classical disk-accreting T Tauri
stars. Current estimates for the frequency and duration of FU Orionis
outbursts and the mass accretion rates of T Tauri and FU Orionis stars
suggest that more mass may be accreted during the outbursts than
during the T Tauri phases. If this is the case, then the outbursts
should also dominate the accretion of angular momentum. During the
outbursts, the accretion rate is so high that the magnetic field of
the star should not disrupt the disk, and the disk will extend all the
way in to the stellar surface. Standard thin disk models then predict
that the star should accrete large amounts of angular momentum, which
will produce a secular spinup of the star.

We present boundary layer solutions for FU Orionis parameters which
show that the rotation rate of the accreting material reaches a
maximum value which is far less than the Keplerian rotation rate at
the stellar surface. This is due to the importance of radial pressure
support at these high mass accretion rates. As a result, the rate at
which the star accretes angular momentum from the disk drops rapidly
as the stellar rotation rate increases. In fact, the angular momentum
accretion rate drops below zero for stellar rotation rates which are
always substantially below breakup, but depend on the mass accretion
rate and on the adopted definition of the stellar radius. When the
angular momentum accretion rate drops close to zero, the star will
stop spinning up. Faster stellar rotation rates will produce negative
angular momentum accretion which will spin the star down. Therefore,
FU Orionis outbursts can keep the stellar rotation rate close to some
equilibrium value for which the angular momentum accretion rate is
small. We show that this equilibrium rotation rate may be similar to
the observed rotation rates of T Tauri stars; thus we propose that FU
Orionis outbursts may be responsible for the observed slow rotation of
T Tauri stars. This mechanism is independent of whether the disk is
disrupted by the stellar magnetic field during the T Tauri phase.

\end{abstract}


\section{Introduction}

The FU Orionis stars are a class of accreting pre-main sequence stars
which are observed to be T Tauri stars undergoing large outbursts (see
Hartmann, Kenyon, \& Hartigan 1993 for a review). During the
outbursts, these systems brighten by $\sim 5-6$ magnitudes over a
period of $\sim 1-10$ years. They reach luminosities of a few hundred
$\lsun$, with spectra indicating maximum temperatures $\sim 6500-7000$
K. The outbursts last for $\sim 100$ years, although this number is
poorly known, since no outbursting systems have been observed to
return to their pre-outburst state. Hartmann \& Kenyon (1985, 1987)
showed that FU Orionis stars contain accretion disks by demonstrating
that spectra of these systems contain double-peaked absorption
lines. Simple steady accretion disk models of two FU Orionis systems,
FU Orionis and V1057 Cygni, were constructed by Kenyon, Hartmann, \&
Hewett (1988, hereafter KHH). The models which best fitted the
observed broad-band spectral energy distributions of these systems had
accretion rates $\dot M \sim 10^{-4} \msyr$, and central stars with
masses $M_* \sim 0.3-1 \msun$ and radii $R_* \sim 4 \rsun$. The
presence of a disk in these systems suggests that their outbursts may
result from a disk instability similar to those believed to cause
similar outbursts in dwarf novae (Clarke, Lin, \& Papaloizou 1989;
Clarke, Lin, \& Pringle 1990; Bell \& Lin 1994; Bell \et 1995).  

Disk accretion also has important implications for the rotational
evolution of the accreting star. In the standard picture of disk
accretion (Shakura \& Sunyaev 1973), the star accretes angular
momentum from the disk. The angular momentum accretion rate is simply
the Keplerian specific angular momentum at the stellar surface
multiplied by the mass accretion rate. This quickly spins up the
accreting star. Thus, in the standard picture, if T Tauri or FU
Orionis stars have accreted even a small fraction of their mass
through a disk, they should be spinning rapidly.  The rotation rates
of FU Orionis stars are not well known, since the accretion luminosity
is so high that it overwhelms the stellar luminosity. The rotation
rates of T Tauri stars are more easily observable and have been
studied in some detail (Bouvier \et 1993, 1995). They tend to be about
a tenth of breakup speed, on average, with rotation periods usually
around 6--8 days. Since the angular momentum added by disk accretion
should rapidly spin T Tauri stars up to breakup speed, some authors
have hypothesized that the spinup of T Tauri stars is controlled by
the interaction between the disk and a strong stellar magnetic field
(K\"onigl 1991; Cameron \& Campbell 1993; Hartmann 1994; Shu \et 1994).

This mechanism for maintaining the spin rate of a star near some
equilibrium value was first explored in detail in order to explain the
spin behavior of some accreting neutron stars (Ghosh, Lamb, \& Pethick
1977; Ghosh \& Lamb 1979a,b). In this model, the stellar magnetic
field disrupts the disk at some distance from the stellar surface. The
location of this disruption radius is determined by the strength of
the magnetic field and the mass accretion rate; a stronger magnetic
field disrupts the disk farther away from the star, while a larger
mass accretion rate moves the disruption radius closer to the stellar
surface. The Keplerian rotation speed at the disruption radius
determines approximately how fast the star will rotate. In order to
produce the rotation speeds of most T Tauri stars, the disk would have
to be disrupted around 5 stellar radii from the stellar surface, and
for the accretion rates commonly ascribed to T Tauri stars, this would
require a stellar magnetic field $\sim$ 1 kG (K\"onigl 1991).

The magnetic hypothesis for explaining the slow rotation of T Tauri
stars does not include the effects of angular momentum accretion
during FU Orionis outbursts. Event statistics of these outbursts, in
which the number of observed events is compared to the number of young
stars in the observable volume, suggest that each star probably
undergoes multiple outbursts (Herbig 1977; Hartmann \& Kenyon
1985). Estimates for the time between outbursts vary widely from $\sim
10^3-10^5$ years (e.g. Bell \& Lin 1994; Kenyon 1995). If the
accretion rate during this period is the typical T Tauri rate of $\sim
10^{-7} \msyr$, then $10^{-4}-10^{-2} \msun$ will be accreted during
the T Tauri phase. During the intervening FU Orionis outbursts, the
accretion rate is $\sim 10^{-4} \msyr$ for $\sim 100$ years, so $\sim
10^{-2} \msun$ should be accreted during the outburst. This suggests
that as an accreting pre-main sequence star evolves through multiple
cycles of FU Orionis outbursts and T Tauri quiescent phases, most of
the mass accreted by the star may be added during the outbursts. If
this is the case, then we should expect that the outbursts will also
dominate the accretion of angular momentum.

During the FU Orionis outbursts, the stellar magnetic field should
have little effect on the accretion disk. As noted above, in order to
produce the observed rotation rates of T Tauri stars, the field should
disrupt the disk at a radius of about 5 stellar radii. When such a
star experiences an FU Orionis outburst, the mass accretion rate
increases by a factor of order 1000. This should be sufficient to
overwhelm the magnetic field, and allow the disk to reach all the way
in to the stellar surface. In the standard picture of accretion disks,
since the star accretes a large amount of mass during the FU Orionis
outburst, it should also accrete a large amount of angular momentum,
which will spin the star up substantially. This will in turn require
that the star loses large amounts of angular momentum during the T
Tauri phase in order to stay at the observed slow rotation rate. If
the star accretes less mass in the T Tauri phase than it does in the
FU Orionis outburst, it will be unable to lose sufficient angular
momentum to spin back down, and over time, the star will continue to
spin up toward breakup speed.

If the disk reaches the stellar surface, and the star is rotating
slowly, a viscous boundary layer forms between the rapidly rotating
disk material and the star. The boundary layer controls the flow of
angular momentum between the disk and the star. Also, if the central
star is not rotating, the boundary layer should produce as much
luminosity as the disk. Clearly, if we wish to understand the angular
momentum transfer and to compare disk models to observed spectra of FU
Orionis systems, it is essential to include the boundary layer in our
disk models.

Our initial investigation of the boundary layer structure examined the
question of whether a star continues to accrete mass and angular
momentum when it has been spun up to breakup speed (Popham \& Narayan
1991, hereafter PN91). We found that as the star approaches the
breakup rotation rate, the amount of angular momentum accreted per
unit of mass accreted drops dramatically, and can even become
negative, so that the star continues to accrete mass while losing
angular momentum to the disk. Thus, there is some equilibrium rotation
rate close to breakup for which the star accretes no angular momentum,
so that it can continue to accrete mass without spinning up or
down. This study used a very simple disk model with a polytropic
relation between the disk pressure and density, so it was not clear
that similar conclusions would hold when a more realistic treatment of
the disk was used.  More recently, we have explored the structure of
the boundary layer region for several types of accreting stars,
including T Tauri and FU Orionis stars (Popham \et 1993, hereafter
PNHK) and cataclysmic variables (Narayan \& Popham 1993; Popham \&
Narayan 1995), using a more realistic model which includes the energy
balance and radiative transfer.

In this paper, we examine the angular momentum transfer between the
star and the disk for boundary layer solutions calculated for
parameters corresponding to FU Orionis systems. The boundary layer
model we use is similar to the one used in our studies of cataclysmic
variables (Narayan \& Popham 1993; Popham \& Narayan 1995) and in our
previous study of pre-main sequence stars (PNHK). We find that many of
the conclusions reached by PN91 continue to apply when our more
detailed model is applied to FU Orionis systems. When the stellar
rotation rate is plotted against the height of the disk at the stellar
surface, the solutions fall on multiple solution branches. We find
that the angular momentum accretion rate drops as the stellar rotation
rate increases, and that solutions exist for a wide range of angular
momentum accretion rates, including negative values. There is an
equilibrium stellar rotation rate for which the star can accrete mass
without spinning up or down. The major difference from the PN91
results is that for these FU Orionis-type solutions, the equilibrium
stellar rotation rate is well below breakup.

On the basis of these solutions, we propose that FU Orionis outbursts
play an important role in the spin evolution of accreting pre-main
sequence stars. The large amount of mass accreted during an outburst
can change the stellar rotation rate substantially. If more mass is
accreted during the FU Orionis outburst phases than during the
intervening T Tauri phases, then the stellar rotation rate will stay
close to the equilibrium rate established during the FU Orionis
phase. Although this equilibrium rate is rather uncertain, we will
show that for reasonable choices of parameters, it is comparable to
the observed rotation rates of T Tauri stars. Thus, angular momentum
transfer during the FU Orionis outbursts can explain the slow rotation
of T Tauri stars without invoking strong stellar magnetic fields. If T
Tauri stars have weak fields which allow the accretion disk to extend
down to the surface of the star and spin up the star, negative angular
momentum accretion during the FU Orionis outburst phases can keep the
star spinning slowly. Alternatively, if T Tauri stars have strong
magnetic fields which disrupt the disk, the FU Orionis phases will
still dominate the spin evolution as long as they dominate the mass
accretion.

In \S 2, we discuss the model used to calculate our boundary layer and
disk solutions for FU Orionis parameters, and the input parameters for
our solutions. We describe a typical solution in detail in \S 3. We
demonstrate the presence of multiple solution branches, and show how
the character of the solutions varies along those branches. We also
show a solution which accretes mass without accreting angular
momentum. Finally, we show how variations in the mass accretion rate
affect the solutions and solution branches. In \S 4, we describe the
choice of an additional condition which allows us to find a relation
between the stellar rotation rate and the angular momentum accretion
rate. We discuss the spin evolution of T Tauri/FU Orionis systems and
the implications of our results in \S 5.

\section{Disk and Boundary Layer Model}

\subsection{Disk Equations}

The model we use to study the boundary layer uses a version of the
``slim disk'' equations developed by Paczy\'nski and collaborators
(Paczy\'nski \& Bisnovatyi-Kogan 1981; Muchotrzeb \& Paczy\'nski 1982).
This model is similar to the models used in previous papers (Narayan
\& Popham 1993; PNHK; Popham \& Narayan 1995), so we will only
describe it qualitatively here, and refer the reader to those papers
for more details.

The model assumes a steady state, so that the mass and angular
momentum accretion rates $\md$ and $\dot J$ are constant with radius.
We use the slim disk versions of the angular momentum, radial
momentum, and energy equations. Thus, we dispense with some of the
simplifying assumptions made in the standard thin disk treatment
(Shakura \& Sunyaev 1973). We include pressure support and
acceleration in the radial momentum equation, so that the rotational
velocity $\O$ is not required to equal the Keplerian value $\ok$. We
allow the angular momentum accretion rate $\dot J$ to deviate from the
standard value $\md \omks \rs^2$ by a factor $j$, so that $\dot J = j
\md \omks \rs^2$. Finally, we drop the assumption that the energy
dissipated by viscosity is radiated locally at the same radius in the
disk. Instead, we include terms for the radial transport of energy by
radiation and by the accreting material itself. The radiative transfer
scheme we use is the one described in PNHK.

The modified angular momentum, radial momentum, and energy equations,
along with the radiative transfer equations, provide a set of
differential equations which can be solved with appropriate boundary
conditions. The most important of these are the conditions on the
angular velocity and the radial radiative flux. The angular velocity
$\O$ must match the stellar rotation rate $\oms$ at $\rs$, and the
Keplerian value $\omko$ at $R=\ro = 100 \rs$. At $\rs$, there is an
outward radiative flux characterized by an effective temperature
$T_*$, which we take to be 5000 K. This produces a luminosity of $\sim
10^{34} \ergs$ entering the boundary layer from the star, which is a
small fraction of the accretion luminosity $L_{acc} \sim 10^{36}
\ergs$. The equations are solved using a relaxation method.

\subsection{Solution Parameters}

Within the framework of the disk and boundary layer model described
above and in the aforementioned papers, a large number of solutions
can be found. These solutions are distinguished by various choices of
input parameters. For our models, these include the mass accretion
rate $\md$, the angular momentum accretion rate $\dot J$, the stellar
mass, radius, and rotation rate $\ms$, $\rs$, and $\oms$, and the
viscosity parameter $\alpha$. In the case of FU Orionis systems, none
of these parameters are directly observable, so they must be estimated
by using the best current models to interpret the available
observations.

\subsubsection{$\md, \ms, and \rs$}

These are the standard accretion disk parameters which determine the
disk luminosity and temperature. Observed spectra of FU Orionis and
V1057 Cygni were fitted with simple accretion disk models by
KHH. Although these models did not include the boundary layer, they
allow us to estimate the basic disk parameters as follows. First, the
infrared flux is produced primarily in regions of the disk which are
fairly far from the stellar surface, so that the flux produced is
insensitive to the value of $\rs$. We have calculated simple blackbody
spectra of our disk and boundary layer solutions for various choices
of $\md$ and $\ms$, and compared them to photometric data of V1057
Cygni. Fits to infrared photometry suggest that the combination $\md
\ms \simeq 5 \times 10^{-5} \msun^2 {~\rm yr^{-1}}$. Our spectra and
fits will be presented in a future paper (Popham \et 1995).

Another constraint is provided by spectra of FU Orionis and V1057
Cygni, which show that the maximum effective temperature reached in
the inner disk is around 7000 K (KHH; Kenyon \et 1989). In the
standard Shakura \& Sunyaev (1973) model, the disk reaches a maximum
effective temperature $T_{max} = 0.287(G \ms \md / \rs^3
\sigma)^{1/4}$ at $R = 49/36 \rs$. $T_{max} \propto \rs^{-3/4}$, so it
provides a reasonable way of estimating $\rs$; however, the
Shakura-Sunyaev formulation does not include any luminosity from the
boundary layer, so the true temperature of the inner disk will be
higher than $T_{max}$. For $\md \ms = 5 \times 10^{-5} \msun^2 {~\rm
yr^{-1}}$, we find $T_{max} \simeq 6500$ K for $\rs = 3 \times 10^{11}
\cm = 4.31 \rsun$. This value for $\rs$ is somewhat larger than the
observational estimates for T Tauri star radii, such as those
tabulated by Bouvier \et (1995), where $\rs \simeq 1.5-2.5 \rsun$.

This combination of parameters gives an accretion luminosity $L_{acc}
\simeq 1.4 \times 10^{36} \ergs \simeq 350 L_\odot$, which agrees with
the bolometric luminosity estimates for most of the FU Orionis systems
listed by Bell \& Lin (1994). In the solutions presented in this
paper, we keep $\ms = 0.5 \msun$ constant, and vary $\md$ and $\rs$ so
that the accretion luminosity remains nearly constant. The three sets
of solutions have $\md = 10^{-4.0}, 10^{-4.15}$, and $10^{-4.3}
\msyr$, and $\rs = 4 \times 10^{11}, 3 \times 10^{11}$, and $2.25
\times 10^{11} \cm$, respectively.

\subsubsection{$\alpha$}

The viscosity parameter $\alpha$ is more difficult to determine. The
most common way to do so is to try to estimate the viscous timescale
in the disk by observing the disk outburst behavior. This assumes that
the FU Orionis outbursts are produced by disk instabilities, or at
least by some disturbance which dies away on the viscous timescale.
Disk instability models of FU Orionis outbursts have found that small
values of $\alpha \sim 10^{-3}-10^{-4}$ were required to reproduce the
long timescales of observed outbursts (Clarke, Lin \& Pringle 1990;
Bell \& Lin (1994); Bell \et 1995). Unfortunately, these values of
$\alpha$ lead to problems with the evolutionary timescales and
gravitational stability of the disk (Bell \et 1995). We have therefore
adopted an intermediate value of $\alpha = 0.01$ which we use for all
of our models.

\subsubsection{$\oms$ and $j$}

These parameters are usually ignored in accretion disk studies, since
in the standard Shakura \& Sunyaev-type model, the accreting star is
assumed to be nonrotating, and the angular momentum accretion rate is
just given by the Keplerian specific angular momentum at the stellar
surface multiplied by the mass accretion rate.  Our study treats the
boundary layer region explicitly, so these assumptions are no longer
necessary. We treat both the stellar rotation rate and the angular
momentum accretion rate as free parameters for purposes of finding our
solutions. Later, when we discuss the spin evolution of accreting
pre-main-sequence stars in \S 4, we will attempt to define a relation
which specifies the angular momentum accretion rate as a function of
the stellar rotation rate.

\section{Solutions and Solution Branches}

\subsection{A Typical Solution}

We begin by discussing an individual boundary layer solution with $\md
= 10^{-4.15} \msyr$ and $\rs = 3 \xec$, with $j=0.90$ and
$H(\rs)=10^{11} \cm$. These parameters are typical of most of the
solutions presented here. Note that we have specified the disk height
at $\rs$ rather than $\oms$; the reason for this is discussed
below. Figure 1 shows the variation of the rotational velocity $\O$,
the radial velocity $v_R$, the effective temperature $\teff$, and the
disk height $H$ in the region $R = 1-3 ~\rs$.

The rotational velocity $\O$ reaches a maximum around $R \simeq 5
\xec \simeq 5/3 ~\rs$, so the dynamical boundary layer is about $2/3
~\rs$ wide. Note that the maximum $\O$ is only about 70\% of
$\Omega_K$ at that radius. This means that even at the outer edge of
the dynamical boundary layer, $\O^2 R$ is only about 50\% of the
gravitational term $\Omega_K^2 R$; the other half of the support
against gravity is provided by the pressure gradient.

The radial velocity $v_R$ shows a similar profile to $\O$. It reaches
a maximum of $\sim 3 \times 10^4 \cm \pers$ at $R \simeq 6 \xec \simeq
2 ~\rs$. Thus, the accreting material moves radially through the
boundary layer region in about 1 year. In the boundary layer, the
radial velocity drops rapidly, reaching $\sim -2000 \cm \pers$ at the
stellar surface.

The effective temperature $\teff$ peaks closer to the stellar surface
than either $\O$ or $v_R$. In fact, the maximum $\teff \sim 8300$
K is reached at $R = 3.8 \xec$, in the middle of the dynamical
boundary layer. The radiative energy released in the dynamical
boundary layer propagates out to larger radii, but most of it has been
radiated from the disk surface inside 2 $\rs$. Thus, the thermal
boundary layer (the region of elevated effective temperature) is only
slightly wider than the dynamical boundary layer.

The disk height $H$ is fairly large: the ratio $H/R$ is close to
1/3 throughout the inner region. Note that in earlier studies of the
boundary layer structure at lower accretion rates, we generally found
that $H$ increased rapidly in the pressure-gradient supported region
inside the dynamical boundary layer. Here, as noted above, the
accreting material is already substantially pressure-gradient
supported at the outer edge of the dynamical boundary layer, and no
rapid increase in $H$ is seen at the innermost radii.

\subsection{Solution Branches}

One goal of this paper is to examine the variation in the boundary
layer structure as the various system parameters discussed in \S 2.2
change. We are particularly interested in the spin history of
accreting pre-main sequence stars, so we focus on the variations
produced by changes in the rotation rate of the star $\oms$ and the
angular momentum accretion rate $j$. Accordingly, we first varied
$\oms$ while keeping all of the other parameters constant, including
$\dot M$, $\ms$, $\rs$, $\alpha$, and $j$. This produced an intriguing
result. We found that for some of our solutions, as we increased
$\oms$, the disk height $\hs \equiv H(\rs)$ increased, while other
solutions showed the opposite behavior: $\hs$ decreased as the star
spun faster.

We continued to systematically increase and decrease $\oms$, and
plotted the values of $\hs$ in the resulting solutions as a function
of $\oms$. This revealed that the tracks traced out in the $\oms -
\hs$ plane connected together into a single curve, like the one shown
in Figure 2. This curve has a Z shape; multiple solutions can be found
for a single value of $\oms$. These solutions have different values of
the disk height $\hs$, and they are distinct solutions, despite the
fact that they share the same values of all of the input parameters
$\md, \ms, \rs, \alpha, \oms,$ and $j$. In fact, by specifying $\hs$
rather than $\oms$, we can find a unique solution. The Z-shaped curve
has three branches; on the middle branch, $\hs$ increases with
increasing $\oms$, while on the upper and lower branches, $\hs$
decreases with increasing $\oms$.  The existence of multiple solution
branches for a single value of $j$ agrees with the findings of our
earlier work (PN91), which used a much simpler disk formulation.

\subsubsection{Variations Along the Branches}

The solutions along a Z-shaped curve show a wide range of variation in
the structure of the boundary layer. In Figure 3 we have plotted 10
solutions with $\hs$ ranging from $8 \xtc$ to $1.25 \xec$. The
positions of these 10 solutions in the $\oms - \hs$ plane are marked
on the Z-shaped curve shown in Fig. 2 The solutions in the upper left
have large values of $\hs$ and small values of $\oms$ corresponding to
a slowly rotating star. These solutions have a very broad boundary
layer, as shown in Figure 3. The peak values of $\O$ and $v_R$ sit out
at $R \simeq 4 \rs$. The peak effective temperature, occurs much
closer to the star, but the region of elevated effective temperature
extends out to several stellar radii. The disk height $H$ is
approximately proportional to $R$, with $H/R \sim 0.4$, as shown in
Fig. 3c.

Solutions with smaller values of $\hs$, but still lying on the upper
branch, have progressively narrower boundary layers. The last solution
shown on the upper branch, with $\hs = 1.05 \xec$, has $\O_{max}$ at
$R < 2 \rs$ (Fig. 3). Meanwhile, $\oms$ increases gradually, reaching
a value of $\sim 3.5 \times 10^{-6} \pers$ at the transition from the
upper to the middle branch. Note that this value of $\oms$ is less
than 10\% of $\omks \simeq 5 \xfs$.

As $\hs$ continues to decrease along the middle branch, the boundary
layer continues to shrink in radial extent. The changes in the
solutions along the middle branch with decreasing $\hs$ seem to
continue smoothly the changes along the upper branch, with one
exception: $\oms$ drops back down toward zero. The $\hs = 9 \xtc$
solution, near the transition from the middle to the lower branch, has
$\O_{max}$ at $R \simeq 4 \xec \simeq 4/3 ~\rs$ (Fig. 3), so the
radial extent of the boundary layer has decreased substantially. This
solution has $\oms \simeq 5 \times 10^{-7} \pers \simeq 0.01 \omks$.

Finally, along the lower branch, the trends noted above for the upper
and middle branches continue. The boundary layer becomes quite narrow,
with $\O_{max}$ at $R \simeq 1.15 \rs$ in the $\hs = 8 \xtc$
solution. The value of $\oms$ increases rapidly, and the solutions
have a large angular velocity gradient $d\O/dR$ at $\rs$. This
concentrated energy release produces a strong peak in $\teff$ near
$\rs$, but $\teff$ begins to drop as $\hs$ continues to decrease,
because the large values of $\oms$ for these solutions means that a
large portion of the rotational energy of the accreting material is
retained rather than being dissipated in the boundary layer. The ratio
$H/R$ varies more with $R$ for the lower branch solutions than for
solutions on the other branches; the $\hs = 8 \xtc$ solution has $H/R
= 0.24$ at $R = \rs$ but $H/R \simeq 1/3$ at $R = 3 \rs$.

\subsection{Angular Momentum Accretion Rate}

Thus far we have kept the angular momentum accretion rate $j$ constant
and showed that a family of solutions with a given value of $j$ traces
out a Z-shaped curve in the $\oms-\hs$ plane. Now we show the effects
of varying the angular momentum accretion rate $j$. For each value of
$j$, and using the same $\md$, $\ms$, $\rs$, and $\alpha$ as before,
we can follow the variations in $\oms$ as we vary $\hs$. These
variations are shown in Figure 4.

As we increase $j$, we find a similar Z-shaped curve in the $\oms -
\hs$ plane, but the curve is displaced to the left, i.e. to smaller
values of $\oms$. For $j=0.91$, the transition between the lower and
middle branches has moved to negative values of $\oms$. The
middle-to-upper branch transition also moves left, but by a smaller
amount. The Z-shaped curves for adjacent values of $j$ nest inside one
another, and they continue to be displaced to smaller $\oms$ as $j$
continues to increase.

If we decrease $j$, the curves are displaced to larger $\oms$, but
they also begin to change shape dramatically. At $j=0.89$, the middle
branch has become much shorter and steeper, so that it extends over a
very narrow range in $\oms$. At $j=0.84$, the middle branch has
vanished completely, leaving a smooth, monotonic curve with $\hs$
decreasing as $\oms$ increases.

If we continue to decrease $j$, we continue to find solutions, even
for $j \leq 0$. In these solutions, the star accretes mass while
losing angular momentum. An example of a $j=0$ solution is shown in
Figure 5. The angular velocity continues to rise all the way in to the
stellar surface. Nonetheless, $\oms$ is far below $\omks$,
so the inner region of the disk is almost completely
pressure-supported. There is no maximum in $\O$, so in essence there
is no boundary layer at all. The disk is fairly thick, with $H/R
\simeq 0.4$. We can find solutions for negative values of $j$, and
even for large negative values such as $j=-10$. As $j$ drops, the disk
continues to thicken. We have included the curves for $j=0$ and $j=-1$
in the $\oms-\hs$ plane in Figure 4. The curves are rather similar to
the $j=0.84$ curve discussed above; $\hs$ decreases monotonically with
increasing $\oms$.

\subsection{Variations with $\md$}

Having described the appearance of the family of curves in the $\oms -
\hs$ plane produced by a range of values of $j$, we can examine the
variations in this family of curves as the mass accretion rate
changes. As discussed in \S 2.2, we also vary $\rs$ along with $\md$
so as to keep the accretion luminosity approximately constant, so that
we examine 3 combinations of $\md$ and $\rs$: $\md = 10^{-4.0},
10^{-4.15}$, and $10^{-4.3} \msyr$, and $\rs = 4 \times 10^{11}, 3
\times 10^{11}$, and $2.25 \times 10^{11} \cm$, respectively.

In Figure 6, we show the families of curves in the $\oms-\hs$ plane
for $\md = 10^{-4.0} \msyr$, $\rs = 4 \xec$ and for $\md = 10^{-4.3}
\msyr$, $\rs = 2.25 \xec$. These can be compared to Fig. 4, which
used intermediate values of $\md = 10^{-4.15} \msyr$, $\rs = 3 \xec$. The shapes of the
curves and their variation with $j$ are quite similar for the 3
different accretion rates. The major difference is that the family of
curves is displaced to larger $\oms$ as $\md$ decreases, and to
smaller $\oms$ as $\md$ increases. Thus, at the lowest accretion rate,
$\md = 10^{-4.3} \msyr$, the transition from Z-shaped to monotonic
curves occurs around $\oms \simeq 1.5 \xfs \simeq 0.2
\omks$. This can be compared to the $\md = 10^{-4.15} \msyr$ case,
where the same transition happens around $\oms \simeq 5 \times
10^{-6} \pers \simeq 0.1 \omks$. At $\md = 10^{-4.0} \msyr$, all of
the Z-shaped curves have moved to $\oms < 0$, corresponding to a star
rotating in the opposite sense to the disk.

\section{Spin Evolution of Pre-Main-Sequence Stars}

\subsection{Angular Momentum Accretion}

In order to study the spin evolution of a disk-accreting star, we need
to know how much angular momentum is accreted by the star as a
function of its rotation rate. The answer to this question will
determine the spin evolution of the star. Of course, the spin
evolution may be complicated by variations in the mass accretion rate
or the mass and radius of the star. These considerations are important
for T Tauri/FU Orionis stars, and we will discuss them later.

To begin with, however, we concentrate on the variation of $j$ as a
function of $\oms$ for a star with constant $\md$, $\ms$, and $\rs$.
The general sense of this variation is clear from Figs. 4 and 6, which
show how $j$ varies with $\oms$ and $\hs$. If we assume for the moment
that $\hs$ stays constant, then $j$ decreases as $\oms$ increases.  At
small values of $\oms$, where the constant-$j$ curves are Z-shaped,
$j$ decreases gradually. As the star spins up, the constant-$j$ curves
become monotonic, and $j$ decreases more rapidly. This suggests that
the spin evolution of the star proceeds as follows: as long as $j$
remains positive, the star continues to spin up. Eventually, $j$
decreases to zero, and the star reaches an equilibrium rotation rate
where it neither spins up nor spins down.

The major uncertainty in this picture is in $\hs$: for a given value
of $\oms$, $j$ also depends strongly on the value of $\hs$. Thus, in
some regions of the $\oms-\hs$ plane, if $\hs$ were to decrease
rapidly enough as $\oms$ increases, $j$ would increase.  As we
discussed in \S 3, we can find solutions by specifying $j$ and either
$\oms$ or $\hs$. In nature, one would expect that only the stellar
rotation rate $\oms$ is an intrinsic parameter of the accreting
system, and that both $j$ and $\hs$ should be determined by the
physics of the accretion flow. In order to reproduce this situation
within our model, we need an additional condition which gives $\hs$ as
a function of $\oms$.  Such a condition will define a track in the
$\oms-\hs$ plane, and the variation of $j$ as a function of $\oms$
along this track provides a basis for understanding the spin evolution
of accreting stars.

\subsection{Need for an Additional Boundary Condition}

The major reason for the uncertainty in $\hs$ is the lack of a clear
definition of the stellar radius $\rs$. In the solutions presented
thus far, we have specified a value for $\rs$, but we have not
specified a boundary condition which ensures that the specified $\rs$
represents the true stellar radius in any physical sense. As a result,
we can obtain a wide range of solutions which have very different
physical characteristics at $\rs$, as shown in Fig. 3. Thus, we need
an additional boundary condition, a definition for $\rs$ which marks
the location of the transition from the boundary layer to the star,
based on the changes in the accretion flow.

This is a difficult task, for several reasons. First, our model is set
up in a way that specifically tries to avoid making distinctions
between the disk, boundary layer and star. All of these regions are
assumed to obey the same equations, with the same physical
parameters. Second, the distinction between star and disk is much more
difficult to make at large accretion rates. This can be seen clearly
in Fig. 1. It is quite difficult to separate the accretion flow into
different regions; $\Omega$, $\teff$ and $H$ all vary smoothly over
large length scales. This can be contrasted with the clear differences
between the disk, boundary layer, and star which are visible in models
for lower accretion rates, like those for cataclysmic variables
(Popham \& Narayan 1995) and T Tauri stars (PNHK). In those models,
$\O$ dropped rapidly in the boundary layer, and then stayed almost
constant in the star; the transition between the two regions was quite 
rapid. Similarly, $H$ began to rise rapidly with decreasing $R$ inside
the boundary layer. In the current solutions, the $\O$ profile shows
only weak indications of leveling off, and $H$ continues to decrease
approximately linearly with $R$. This makes it quite difficult to
select a radius which represents a transition from boundary layer to
star. 

Another perspective on this issue comes from examining the range of
solutions in Fig. 3, all of which use a single value of $j$. One way
to locate the stellar radius is to select a solution from this set for
which the boundary layer--star transition appears to occur at the
inner edge of our computational grid, where $R = \rs$. It seems clear that the solutions with the smallest values of
$\hs$ are not appropriate. They have very large values of $d\O/dR$ at
$\rs$, and it seems clear that if the solution were to continue in to
$R < \rs$, $\O$ would continue to drop precipitously. Thus, it appears
that in these solutions, $\rs$ falls in the middle of the boundary
layer, and the boundary layer--star transition would occur at $R <
\rs$. On the other hand, the solutions with the largest values of
$\hs$ have very broad regions, extending out to a few times $\rs$, in
which $\O$ decreases quite gradually. It appears that these solutions
have reached the boundary layer--star transition at $R > \rs$, and the
material then simply continues to settle as it flows in slowly to
$\rs$. The intermediate solutions with $\hs \simeq 10^{11} \cm$ seem
to have the best characteristics. Both $\O$ and $v_R$ have dropped
substantially but are leveling off around $\rs$.

\subsection{Choice of a Boundary Condition}

We have many possible choices for boundary conditions to specify
$\rs$. One possibility would be to try to match the shape of the disk
surface to the shape of the surface of a star rotating at a speed
$\oms$; however, as we have discussed, the disk height $H$ does not
vary in a way that resembles the transition from a disk to a star, so
this would be difficult. The best variables to use for a boundary
condition are probably the angular and radial velocities $\O$ and
$v_R$. These variables show directly where the boundary layer is
located, and compared to other variables like the central and
effective temperatures of the disk, they depend far less on the
details of radiative transfer, opacities, etc.

Of the two velocities, we believe that the radial velocity 
provides a superior measure of the location of the boundary
layer--star transition. This is because the angular velocity may or
may not drop as the accreting material approaches the stellar surface;
if the star is rapidly rotating, $\O$ may drop only slightly, or may
continue to increase, as we saw in the solutions with small and
negative values of $j$ in \S 3.3. In such cases, it becomes very
difficult to distinguish the disk from the star on the basis of the
$\O$ profile. The radial velocity, on the other hand, always reaches a
maximum value and then drops substantially as it approaches the
stellar surface. Put another way, $\oms$ does not have to be small,
but $\vs$ does.

The simplest condition we can put on $v_R$ is simply to select a value
for $\vs$, and define $\rs$ as the radius where $v_R = \vs$. This can
then be used to determine the variation of $\hs$ and $j$ with
$\oms$. One simple way to do this is to keep $j$ constant and vary
$\hs$, so as to move along one of the constant-$j$ curves shown in
Figs. 4 and 6. Fig. 3 shows that $v_R$ varies monotonically along a
constant-$j$ curve, such that $v_R$ decreases with increasing
$\hs$. Thus, by varying $\hs$, we can find a solution for which $v_R =
\vs$ at $R = \rs$. Corresponding solutions can be found for other
values of $j$, and the values of $\oms$ and $\hs$ for these solutions
will then produce the desired relations for $\hs(\oms)$ and $j(\oms)$.

\subsection{Variation of $j$ with $\oms$}

In order to illustrate how $j$ varies with $\oms$ for this boundary
condition, we choose a typical value of $\vs = 1000 \cm \pers$. This
value is a factor of $\sim 2 \times 10^4$ smaller than the free-fall
velocity, and it is 5-15\% of the maximum radial velocity reached in
the resulting solutions. The variation of $\hs(\oms)$ for the
solutions with $v_R = \vs$ at $\rs = 3 \xec$ is shown as a
dashed line in Fig. 4. Note that $\hs$ is nearly constant for the full
range of $\oms$, with $\hs \simeq 1.07 - 1.11 \xec$, so that $\hs / \rs
\simeq 0.36$. There is some variation of $\hs$; it decreases gradually
from about $1.088 \xec$ at $\oms = 0$ to $1.075 \xec$ at $\oms
\simeq 1.3 \xfs$, and then rises more rapidly, reaching about $1.108
\xec$ at $\oms \simeq 1.83 \xfs$.

More importantly, $j$ decreases with increasing $\oms$; as the star
spins up, it accretes less angular momentum. Figure 7a shows the
variation of $j$ as a function of $\oms$; $j$ decreases gradually at
first, and then drops more rapidly. The angular momentum accretion
rate reaches zero at $\oms \simeq 1.5 \xfs$, and rapidly drops to $j=
-1$ at $\oms \simeq 1.8 \xfs$. Note that these values of $\oms$ are
only about one third of the Keplerian angular velocity at this radius,
$\omks \simeq 5 \xfs$. Thus, the star stops accreting angular momentum
when it is still rotating well below breakup.

At other mass accretion rates, the variation of $\hs$ and $j$ with
increasing $\oms$ is quite similar: $\hs$ stays approximately constant
and $j$ decreases gradually at first, then more rapidly. The major
difference is that $j$ decreases more rapidly at higher mass accretion
rates, and reaches $j=0$ at a smaller value of $\oms$. This is
illustrated in Figure 7b and c, where we plot $j(\oms)$ for the $\md =
10^{-4.0} \msyr$ and $\md = 10^{-4.3} \msyr$ solutions. We have used
the same criterion, $\vs = -1000 \cm \pers$, to define the stellar
radius. At $\md = 10^{-4.0} \msyr$, $j=0$ for $\oms \simeq 6.6 \xss$,
corresponding to a rotation period of $\sim 11$ days. Also note that
for this higher accretion rate, $j$ is substantially smaller than one
even at $\oms = 0$. At $\md = 10^{-4.3} \msyr$, $\oms(j=0) \simeq 3
\xfs$, which corresponds to a much shorter rotation period of $\sim
2.4$ days. Recall that as $\md$ changes, we have also varied $\rs$ by
a similar amount in order to keep the accretion luminosity
constant. Thus the variation in $\oms(j=0)$ as a fraction of the
breakup rotation rate $\omks$ is not as dramatic as the variation in
$\oms(j=0)$ itself; we find $\oms(j=0) / \omks \simeq 0.2, 0.3, 0.4$
for $\md = 10^{-4.0}, 10^{-4.15}, 10^{-4.3} \msyr$, respectively.

\section{Discussion}

\subsection{Spin Evolution of T Tauri/FU Orionis Stars}

We have found that boundary layer solutions for parameters
corresponding to FU Orionis outbursts have small and negative angular
momentum accretion rates for fairly low stellar rotation rates. This
suggests that angular momentum transfer during the FU Orionis
outbursts can keep the rotation rate of T Tauri/FU Orionis stars near
to an equilibrium rotation rate $\omseqf \simeq \oms(j=0)$. If the
star is rotating faster than $\omseqf$ when it experiences an FU
Orionis outburst, it will lose angular momentum in the outburst phase
and spin back down to $\omseqf$. If, on the other hand, the star is
rotating slowly when an outburst occurs, it will have $j \sim 1$ in
the outburst phase and will spin up toward $\omseqf$. We have found
that for accretion rates $\md = 5 \times 10^{-5} - 10^{-4} \msyr$,
$j=0$ for spin rates $\oms(j=0) \simeq 0.66-3.04 \xfs \simeq \omseqf$,
corresponding to rotation periods of $2.4-11$ days. These are similar
to the observed rotation periods of T Tauri stars (Bouvier \et 1993,
1995); however, $\oms(j=0)$ is sensitive to the value assumed for
$\vs$, as discussed below.

During the T Tauri phase, the stellar rotation rate will tend to move
toward a different equilibrium value $\omseqt$. If the stellar
magnetic field is too weak to disrupt the disk, then the star will
spin up with $j \sim 1$ until it reaches $\omseqt \simeq \omks$. If
the stellar field does disrupt the disk, then $\omseqt$ will be
determined by the radius where the disruption occurs, and may be
smaller or larger than $\omseqf$. Therefore $j$ may be positive or
negative, with a magnitude $|j| ~\sles~ (R_d/\rs)^{1/2}$ (Ghosh, Lamb,
\& Pethick 1977), where $R_d$ is the radius at which the field
disrupts the disk. Models of magnetically disrupted disks in T Tauri
stars suggest $R_d \sim 4-5 \rs$, which would give $j \sim \pm 2$.

The spin evolution of a T Tauri/FU Orionis star will depend on
$\omseqf$, $\omseqt$, and on the amounts of mass accreted in typical
FU Orionis and T Tauri phases, $\dmf$ and $\dmt$.  If $\dmf > \dmt$,
as suggested by current estimates of the frequency and duration of FU
Orionis outbursts, the stellar rotation rate $\oms$ will move toward
$\omseqt$ during each T Tauri phase, but will return to $\omseqf$
during each FU Orionis phase.

How large will the excursions in $\oms$ be? A rigidly rotating star
has angular momentum $J = I \oms$, where $I$ is the star's moment of
inertia, so $\dot J = \dot I \oms + I \dot \oms$. If we assume that
the moment of inertia changes slowly, so that $\dot J \simeq I \dot
\oms$, then the star spins up and down at a rate $\dot \oms / \omks =
(j/k)(\md / M)$, where we have defined $I \equiv k M \rs^2$. Since $k$
is substantially less than 1, the increase in $\oms$ as a fraction of
the breakup rotation rate $\omks$ is several times larger than the
fractional increase in the stellar mass $M$. If we assume $j=1$,
$k=0.2$, then the accretion of $\Delta M = 0.01 \msun$ onto a $0.5
\msun$ star would increase $\oms$ by $0.1 \omks$. If the typical
change in $\oms$ during the T Tauri phase is larger than the
difference between the equilibrium spin rates $\omseqf$ and $\omseqt$,
then $\oms$ will reach $\omseqt$ during each T Tauri phase, and remain
there for the duration of the phase.

It is important to note that even though the FU Orionis phases may
produce more mass accretion than the T Tauri phases, $\dmf > \dmt$, a
T Tauri/FU Orionis star will still spend most of its lifetime with
$\oms$ moving away from $\omseqf$, due to the short duration of the FU
Orionis phases. This means that the mean value of $\oms$ during the T
Tauri phase will be offset somewhat from $\omseqf$. The sign of this
offset will depend on whether the star spins up or down during the T
Tauri phase, i.e. on whether $\omseqt$ is smaller or larger than
$\omseqf$. Also, the return to the T Tauri state after an FU Orionis
outburst may be gradual, with the mass accretion rate declining slowly
over a period of a few decades. At mass accretion rates intermediate
between T Tauri and FU Orionis rates, the star should spin up, since
the magnetic field may still be too weak to disrupt the disk, while
the mechanism described in this paper will only operate at a large
value of the equilibrium rotation rate. Despite these deviations, the
high accretion rate of mass and angular momentum in the FU Orionis
phase will return $\oms$ to $\omseqf$ quite rapidly, and as long as
$\dmf > \dmt$, $\oms$ will continue to be close to $\omseqf$.

Note also that in general the equilibrium spin rate $\omseq$ will
differ slightly from the value of $\oms$ which gives $j=0$.  Since
$\dot J = I \dot \oms + \dot I \oms$, we will have $\dot \oms = 0$ for
$\dot J = \dot I \oms$. This simply means that in equilibrium, $\dot
J$ must be sufficient to compensate for changes in the star's moment
of inertia. If the moment of inertia increases in response to
accretion, $j > 0$ at $\omseq$. If the moment of inertia decreases, as
it might in an accreting white dwarf, then $j(\omseq) < 0$. For
instance, if the star's radius of gyration stays constant, so that
$\dot I / I = \md / M$, then $j(\omseq) = k \omseq / \omks$, where
again $I \equiv k M \rs^2$. If $\omseq / \omks$ is in the range
0.2-0.4 as suggested by our results, and $k \simeq 0.2$, then
$j(\omseq) \simeq 0.04-0.08$, and $\omseq$ will be only slightly
smaller than $\oms(j=0)$.

\subsection{Solutions with Negative Angular Momentum Accretion}

One of the most important results of this paper is that it establishes
the existence of disk and boundary layer solutions with small and
negative angular momentum accretion rates. In these solutions, the
central star accretes mass but loses angular momentum. This is quite
different from the standard thin disk formulation, in which the
angular momentum accretion rate is always $\dot J = \md \omks \rs^2$,
or $j=1$ in our units. The key to understanding this result is that
the $\O$ profiles of our solutions can be very different from the
purely Keplerian $\O$ assumed in the thin disk case. Most importantly,
$\O$ does not reach a maximum and then drop down to the stellar
rotation rate $\oms$. Instead, $\O$ continues to increase all the way
in to the stellar surface at $R=\rs$. The fact that $\O$ is monotonic
means that there is no radius $R_{\O_max}$ at which $d\O/dR = 0$,
where the viscous torque would vanish, and the angular momentum
accretion rate would simply be the amount of angular momentum carried
in the accreting material at that radius, $\dot J = \md \O_{max}
R_{\O_max}^2$. Since for a steady flow $\dot J$ must be constant with
radius, this would constrain $\dot J$ to be positive. In our low-$j$
and negative-$j$ solutions, there is no maximum in $\O$, so that the
viscous torque carries angular momentum outward at all radii.  If the
torque exceeds the rate at which angular momentum is carried in by the
accreting material, the net angular momentum accretion rate will be
negative.

These solutions are very similar to the ones we found in an earlier
paper (PN91). In that paper, we calculated the structure of the
boundary layer using a simplified disk model with a polytropic
pressure-density relation. We found that our solutions formed two
branches in the $\oms - \hs$ plane for large values of $j \sim 1$. We
did not find the lower branch, but seems likely that it exists for
these solutions as well. The PN91 solution branches formed sharp
corners where they met, unlike the rounded transitions between
branches seen in the current solutions. This difference probably
arises from the very different disk thicknesses in the two studies;
the PN91 solutions generally had $H/R < 0.01$. We were also able to
find solutions with small and negative values of $j$; for these values
only a single branch was present in the $\oms - \hs$ plane. Thus,
despite the use of a different disk model, the qualitative behavior of
these solutions is very close to that of the solutions described in
this paper.

One important question is why the solutions presented here reach
negative angular momentum accretion rates when the stellar rotation
rate $\oms$ is still well below the breakup rate $\omks$. For the PN91
solutions, we defined the stellar radius as the point where $H/R =
0.1$ in the accretion flow, and found that with this definition, $j$
was very close to 1 for most $\oms$, but dropped precipitously when
$\oms \simeq 0.915 \omks$. We have seen that the current solutions
reach $j=0$ for $\oms \sim 0.2 - 0.5 \omks$. As discussed above,
solutions with small or negative values of $j$ must have $d\O/dR < 0$
for all $R$. Thus, if $\O \simeq \ok$ close to $\rs$, the star must
be rotating close to breakup, $\oms \simeq \omks$, in order to allow a
negative-$j$ solution. If $\oms$ is much less than $\omks$, $\O$ will
have to decrease close to $\rs$, and there will be a maximum in $\O$,
so that $j \simeq 1$.

In the FU Orionis solutions presented here, two factors combine to
produce small and negative values of $j$ when $\oms \ll \omks$. The
first is the large radial width of the boundary layer. In the
solutions with $j \sim 1$, $\O$ peaks fairly far from $\rs$, so that
the radial width of the boundary layer is comparable to $\rs$. At the
radius where $\O$ peaks, $\ok$ is substantially smaller than $\omks$;
for instance, at $R = 2 \rs$, $\ok \simeq 0.35 \omks$. As the star
spins up, $\oms$ only needs to reach this value to make $d\O/dR < 0$
everywhere and allow $j$ to be small or negative. Second, pressure
support plays an important role, and $\O$ is substantially smaller
than $\ok$. This allows solutions with $d\O/dR < 0$ everywhere for
even smaller values of $\oms$. This also explains why we reach $j=0$
at smaller $\oms$ for larger values of $\md$; as $\md$ increases,
pressure support plays a larger role, the disk becomes thicker, and
$\O$ is a smaller fraction of $\ok$.

It is also worth noting that the total accretion luminosity increases
substantially as $j$ decreases. In general, the accretion luminosity
varies according to the expression \[ L_{acc} \simeq {G \ms \md \over
\rs} \left[ 1 - j {\oms \over \omks} + {1 \over 2} {\oms^2 \over
\omks^2} \right] \] (Popham \& Narayan 1995). Thus, for large negative
values of $j$, the accretion luminosity can be substantially larger
than the standard value $G \ms \md / \rs$. If the star spins up
substantially during the T Tauri phase, then when an FU Orionis
outbursts occurs, $j$ may reach fairly large negative values. For
instance, Fig. 4 shows that $j = -1$ for $\oms \simeq 1.83 \xfs$,
which is only about a 6\% increase in $\oms/\omks$ over the point
where $j=0$. A 10\% variation from $\oms/\omks \simeq 0.3$ to 0.4, as
discussed above, would give $j \sim -2$. This would give an accretion
luminosity of $\sim 1.88 G \ms \md / \rs$, which would make the
outburst even brighter than one would expect from the increase in
$\md$. This additional luminosity would come from the rotational
energy of the star, which is released as the star spins down.

Disk solutions with negative angular momentum accretion rates may be
important in other types of accreting systems. In systems such as
cataclysmic variables, the accretion rates are generally much smaller
than those of FU Orionis systems. This produces a narrower boundary
layer, and pressure support plays a smaller role, so we would expect
that negative angular momentum accretion rates are only found for
$\oms \simeq \omks$. We are currently completing a study of
negative-$j$ solutions for disks around cataclysmic variables with
$\md = 10^{-8} \msyr$ which confirms these expectations (Popham
1995). In some other systems, such as embedded pre-main-sequence
stars, or some X-ray binaries, the accretion rates may be large enough
to produce solutions like the ones presented here, where negative
angular momentum accretion rates are reached when the star is still
spinning relatively slowly.

\subsection{Assumptions}

The results described above depend on a number of assumptions. One
very important assumption is that FU Orionis outbursts occur in
classical T Tauri stars; i.e., that the T Tauri and FU Orionis phases
occur during the same epoch of pre-main sequence stellar
evolution. Models of FU Orionis outbursts based on disk instabilities
generally assume that the outbursts arise in T Tauri star disks, so
that the FU Orionis and T Tauri phases correspond to the high and low
states observed in other types of accreting stars, such as dwarf
novae. The observational connection between T Tauri and FU Orionis
stars is rather tenuous, since only one of the current FU Orionis
systems was observed spectroscopically before it went into outburst,
and no known FU Orionis systems have yet returned to their
pre-outburst levels. The one system for which a spectrum was taken
prior to outburst, V1057 Cygni, resembled a typical T Tauri star
(Herbig 1977). However, some of the observed characteristics of FU
Orionis systems, such as their association with reflection nebulae and
outflows, and their large far-infrared excesses, suggest that FU
Orionis outbursts may arise in younger embedded sources rather than in
classical T Tauri stars (Kenyon 1995). If this is the case, then the
mechanism described here should serve to limit the stellar rotation
rate during the embedded phase of pre-main-sequence stellar
evolution. The rotation rate would then presumably be controlled by
the stellar magnetic field during the T Tauri phase.

We must also make some assumption about the location of the stellar
radius, which is difficult to define in our model (see \S 4 for a
discussion of this problem). We have chosen a particular value of
$\vs$ to define $\rs$, in order to provide an example of the variation
of $\hs$ and $j$ as a function of $\oms$. Unfortunately, the value of
$\oms(j=0)$ is fairly sensitive to the choice of $\vs$. We have
illustrated this in Fig. 4, where we have plotted lines corresponding
to $\vs = -500 \cm \pers$ and $\vs = -2000 \cm \pers$. For these
values of $\vs$, $j$ reaches zero at $\oms \simeq 0.94$ and $2.19
\xfs$, respectively. This means that for a factor of 4 variation in
$\vs$, $\oms(j=0)$ varies by more than a factor of 2. Thus, even when
we specify all the parameters of an FU Orionis system, we still have a
substantial uncertainty in $\oms(j=0)$ resulting from the uncertainty
in the definition of the stellar radius. As discussed in \S 4.5,
$\oms(j=0)$ also varies dramatically with $\md$. Therefore we can only
say that for what we believe to be reasonable choices of $\vs$ and
$\md$, the resulting values of $\oms(j=0)$ are comparable to T Tauri
rotation rates.

Our results are not strongly dependent on the magnetic field of the
star, as long as the field is not strong enough to disrupt the
accretion flow in the high-$\md$ FU Orionis phase. This would require
a field substantially larger than the $\sim 1$ kG needed to disrupt
the disk during the T Tauri phase. The disruption radius varies as
$R_d \propto \md^{-2/7} B^{4/7}$ (Pringle \& Rees 1972), so to
maintain the same $R_d$ requires that $B$ vary as $\md^{1/2}$. Since
FU Orionis accretion rates are a factor of $\sim 1000$ higher than T
Tauri rates, $B$ would have to be about 30 times larger to disrupt the
disk, which would require $B \sim 3 \times 10^4$ G. A field this
strong would disrupt T Tauri disks at $R ~\sgreat~ 20 \rs$, which can
probably be ruled out by observations.

We have used stellar radii of 2,25, 3, and 4 $\xec$, corresponding to
2.88, 4.31, and 5.75 $\rsun$, in our calculations. These are larger
than the radii generally found for T Tauri stars, which are in the
range 1.5--2.5 $\rsun$. These larger radii appear to be required by
observations: smaller stellar radii would produce maximum temperatures
in the inner disk which are substantially larger than those
observed. This suggests that the accreting star may expand during the
FU Orionis outburst.

Prialnik \& Livio (1985) calculated the evolution of a fully
convective 0.2 $\msun$ star during the accretion of $2.5 \times
10^{-3} \msun$ of material at rates ranging from $10^{-1}$ to
$10^{-10} \msyr$. They also varied the fraction $f$ of the accretion
energy carried into the star with the accreting material; $f$ ranged
from 0.001 to 0.5. They found that for the accretion rates appropriate
to FU Orionis outbursts, the star expanded stably if $f \leq 0.03$. If
$f \geq 0.1$, the star expanded unstably. In our FU Orionis solutions,
the central temperature of the disk at $\rs$ is $\sim 1-3 \times 10^5$
K. The virial temperature for these parameters is $\sim 10^6$ K, so a
reasonable fraction $f \sim 0.1-0.3$ of the accretion energy is
advected into the star. This suggests that rapid stellar expansion may
result from the high-$\md$ accretion during an FU Orionis outburst,
but it is difficult to know how the mass of the star, the presence of
rotation, and the departure from spherical symmetry would affect the
Prialnik \& Livio results. This expansion, and the subsequent
contraction during the T Tauri phase, could affect the spin evolution
of the star, and other aspects of the star's interaction with the
disk, and clearly deserves further study.

Our models also assume that the disk and boundary layer are in a
steady state. This is clearly not true for FU Orionis systems, since
they are experiencing outbursts. The viscous timescale is $t_{visc}
\simeq R^2/\nu = \alpha^{-1} (H/R)^{-2} \ok^{-1}$. For our solutions
with $\alpha = 0.01$, $H/R \simeq 0.3$, this gives $t_{visc} \simeq
1000 \ok^{-1}$. In the inner disk, we have $\ok \simeq {\rm few}
\xfs$, so $t_{visc} \simeq {\rm few} \times 10^7 ~{\rm s} \simeq 1
~{\rm yr}$. This is comparable to the rise times of the fastest-rising
systems, FU Ori and V1057 Cyg, and much shorter than their decline
times. Thus FU Orionis systems should be reasonably well-modeled by
steady state solutions for this value of $\alpha$.

\subsection{Comparison with Observations}

Recent measurements of periodic photometric variations in T Tauri
stars have clearly established that the classical disk-accreting T
Tauri stars spin more slowly than weak-line T Tauri stars, which
appear to lack accretion disks (Bouvier \et 1993, 1995; Edwards \et
1993). This has generally been interpreted to mean that the spinup of
T Tauri stars is controlled by a stellar magnetic field strong enough
to disrupt the disk. If FU Orionis outbursts regulate the spin
evolution of T Tauri stars in the way described in this paper, it
would also account for this observation. In both cases, disk accretion
maintains the slow rotation speed of classical T Tauri stars. If FU
Orionis outbursts cease late in the classical T Tauri stage, then the
more rapid rotation of the weak-line T Tauri stars could be produced
by accretion at the end of the classical T Tauri stage.

Observations of the spin evolution of T Tauri stars could be used to
test the ideas discussed in this paper. The most useful data would be
measurements of the spin period before and after an FU Orionis
outburst. If the star spins down during an outburst, it would offer
clear evidence that outbursts are regulating the rotation rates of T
Tauri stars. If the star spins up, the result is less clear. If the
amount of spinup is less than expected based on the amount of mass
accreted during the outburst, it would suggest that the star is
reaching the equilibrium rotation rate $\omseqf$ at some point during
the outburst. In this case the outbursts would still be limiting the
rotation rate of the star. If, on the other hand, the star spins up as
much as expected, it would suggest that the rotation rate must be
limited by processes taking place during the T Tauri phase. One
complication in interpreting the observations is the possibility that
significant amounts of angular momentum might be carried off in the
strong outflows which occur during outbursts.

Observations of variations in the rotation rates of T Tauri stars
which have not experienced observed outbursts could also be useful.
Over a short timescale, variations in $\md$ could produce irregular
period variations. Nonetheless, if FU Orionis outbursts are
controlling the rotation rates of these stars, then their spin periods
should vary secularly during the T Tauri phase as the rotation rate
$\oms$ moves from $\omseqf$ toward $\omseqt$. If instead the rotation
rates of T Tauri stars remain constant over a long timescale, it would
suggest that $\oms$ has stabilized at $\omseqt$, which would indicate
that the spin evolution is being controlled by angular momentum
transfer in the T Tauri phase.

The main sources of data which can be used to further constrain the
solutions presented in this paper are the spectra of FU Orionis
systems. These spectra yield valuable information on the disk
temperatures and rotational velocities (Hartmann \& Kenyon 1985, 1987;
KHH). The solutions presented in this paper were selected to have
luminosities and temperatures in approximate agreement with those
derived from observations. Because our main aim in this paper has been
to understand the spinup and angular momentum transfer in FU Orionis
systems, we have not attempted to match our solutions with
observations in any detail. We defer this to a subsequent paper
(Popham \et 1995), where we plan to compare the spectra and rotational
velocities produced by our disk and boundary layer solutions to those
observed in FU Orionis systems.

\acknowledgements

I would like to thank Ramesh Narayan, Scott Kenyon, and Lee Hartmann
for helpful discussions. I would also like to thank the organizers of
the conference on Circumstellar Disks, Outflows and Star Formation,
held in Cozumel, Mexico, where an early version of this work was
presented. This work was supported by NASA grant NAG5-2837 at
the Center for Astrophysics and grants NASA NAGW-1583, NSF AST
93-15133, and NSF PHY 91-00283 at the University of Illinois.

\newpage

\figcaption{A boundary layer solution for typical FU Orionis
parameters. The solution shown here has $\md = 10^{-4.15} \msyr$, $\ms
= 0.5 \msun$, $\rs = 3 \xec$, $j = 0.90$, $\hs = 10^{11} \cm$, and
$\alpha = 10^{-2}$. The four panels show the angular velocity $\O$,
the radial velocity $v_R$, the effective temperature $\teff$, and the
disk height $H$ in the range from $R = 3-9 \xec = 1-3 \rs$. The dashed
line shows the Keplerian angular velocity $\ok$.}

\figcaption{Variation of the rotation rate $\oms$ and the disk height
$\hs$ at $R = \rs = 3 \xec$, for solutions with $j=0.90$, $\md =
10^{-4.15} \msyr$, $\ms = 0.5 \msun$, $\rs = 3 \xec$, and $\alpha =
10^{-2}$. The solutions form a Z-shaped curve in the $\oms - \hs$
plane, with three solution branches.}

\figcaption{Same as Fig. 1, but for the 10 solutions marked along the
Z-shaped curve in Fig. 2, with $\hs = 0.80, 0.85, 0.90, ... 1.25
\xec$. The boundary layer width increases rapidly as $\hs$ increases,
and angular velocity, radial velocity, and effective temperature all
decrease.}

\figcaption{Same as Fig. 2, but for various values of $j$ as
labeled. All of the solutions on all curves have $\md = 10^{-4.15}
\msyr$, $\ms = 0.5 \msun$, $\rs = 3 \xec$, and $\alpha = 10^{-2}$. As
$\oms$ and $\hs$ increase, $j$ decreases, and the curves change from
Z-shaped to monotonic. The dotted line marks the solutions with $\vs =
-1000 \cm \pers$, the upper dashed line $\vs = -500 \cm \pers$, the
lower dashed line $\vs = -2000 \cm \pers$.}

\figcaption{A boundary layer solution with $j=0$, where the star
accretes mass but no angular momentum. This solution has $\md =
10^{-4.15} \msyr$, $\ms = 0.5 \msun$, $\rs = 3 \xec$, $\hs
\simeq 1.08 \xec$, and $\alpha = 10^{-2}$. Note that $\vs = -1000 \cm
\pers$, as discussed in \S 4.4; if this condition is used to define
$\rs$, then $j=0$ for $\oms \simeq 1.55 \xfs$, corresponding to a
rotation period of 4.7 days.}

\figcaption{Same as Fig. 4, but for two other choices of $\md$ and
$\rs$: (a) has $\md = 10^{-4.0} \msyr$ and $\rs = 4 \xec = 5.75
\rsun$, and (b) has $\md = 10^{-4.3} \msyr$ and $\rs = 2.25 \xec =
3.23 \rsun$. Note that as $\md$ increases, the Z-shape of the curves
becomes less pronounced and only appears at lower values of
$\oms$. The dotted lines mark the solutions with $\vs = -1000 \cm
\pers$.}

\figcaption{Variation of the angular momentum accretion rate $j$ with
the stellar rotation rate $\oms$, for a set of solutions with a given
value of $v_R$ at $\rs$: (a) is for $\md = 10^{-4.15} \msyr$, $\rs = 3
\xec$, and shows $j(\oms)$ for $\vs =$ -500, -1000, and -2000 $\cm
\pers$; (b) is for $\md = 10^{-4.0} \msyr$, $\rs = 4 \xec$, and $\vs
=-1000 \cm \pers$; and (c) is for $\md = 10^{-4.3} \msyr$, $\rs = 2.25
\xec$, and $\vs=-1000 \cm \pers$. In all cases, $j$ drops with
increasing $\oms$ and reaches $j=0$ for $\oms$ well below $\omks$.}


\begin{references}

\reference {} Bell, K. R., \& Lin, D. N. C. 1994, \apj, 427, 987.

\reference {} Bell, K. R., Lin, D. N. C., Hartmann, L. W., \& Kenyon,
S. J. 1995, \apj, 444, 376.

\reference {} Bouvier, J., Cabrit, S., Fernandez, M., Martin, E. L.,
\& Matthews, J. M. 1993, \aap, 272, 176.

\reference {} Bouvier, J., Covino, E., Kovo, O., Martin, E. L.,
Matthews, J. M., Terranegra, L., \& Beck, S. C. 1995, \aap, 299, 89.

\reference {} Cameron, A. C., \& Campbell, C. G. 1993, \aap, 274, 309.

\reference {} Clarke, C. J., Lin, D. N. C., \& Papaloizou,
J. C. B. 1989, \mnras, 236, 495.

\reference {} Clarke, C. J., Lin, D. N. C., \& Pringle, J. E. 1989,
\mnras, 242, 439. 

\reference {} Edwards, S., \et 1993, \aj, 106, 372.

\reference {} Ghosh, P., Lamb, F. K., \& Pethick, C. J. 1977, \apj,
217, 578. 

\reference {} Ghosh, P., \& Lamb, F. K. 1979a, \apj, 232, 259. 

\reference {} Ghosh, P., \& Lamb, F. K. 1979b, \apj, 234, 296. 

\reference {} Hartmann, L. 1994, in Theory of Accretion Disks 2,
ed. W. J. Duschl \et (Dordrecht: Kluwer), p. 19.

\reference {} Hartmann, L., \& Kenyon, S. J. 1985, \apj, 299, 462.

\reference {} Hartmann, L., \& Kenyon, S. J. 1987, \apj, 312, 243.

\reference {} Hartmann, L., Kenyon, S. J., \& Hartigan, P. 1993, in
Protostars and Planets III, ed. E. H. Levy \& J. I. Lunine (Tucson:
Univ. of Arizona Press), 497.

\reference {} Herbig, G. H. 1977, \apj, 217, 693.

\reference {} Hubeny, I. 1990, \apj, 351, 632.

\reference {} Kenyon, S. J. 1995, Rev. Mex. A. A. (Serie de
Conferencias), 1, 237.

\reference {} Kenyon, S. J., Hartmann, L., \& Hewett, R. 1988, \apj,
325, 231 (KHH).

\reference {} Kenyon, S. J., Hartmann, L., Imhoff, C. L., \&
Cassatella, A. 1989, \apj, 344, 925.

\reference {} K\"onigl, A. 1991, \apj, 370, L39.

\reference {} Muchotrzeb, B., \& Paczy\'nski, B. 1982, Acta Astron.,
32, 1.

\reference {} Narayan, R., \& Popham, R. 1993, \nat, 362, 820.

\reference {} Paczy\'nski, B., \& Bisnovatyi-Kogan, B. 1981, Acta
Astron., 31, 283.

\reference {} Popham, R., \& Narayan, R. 1991, \apj, 370, 604 (PN91).

\reference {} Popham, R. 1995, in preparation.

\reference {} Popham, R., \& Narayan, R. 1995, \apj, 442, 337.

\reference {} Popham, R., Narayan, R., Hartmann, L., \& Kenyon,
S. 1993, \apjl, 415, L127 (PNHK).

\reference {} Popham, R., Narayan, R., Hartmann, L., \& Kenyon,
S. 1995, in preparation.

\reference {} Prialnik, D., \& Livio, M. 1985, \mnras, 216, 37.

\reference {} Pringle, J. E., \& Rees, M. J. 1972, \aap, 21, 1.

\reference {} Shakura, N. I., \& Sunyaev, R. A. 1973, \aap, 24, 337.

\reference {} Shu, F. H., Najita, J., Ostriker, E., Wilkin, F., Ruden,
S., \& Lizano, S. 1994, ApJ, 429, 781.

\end{references}
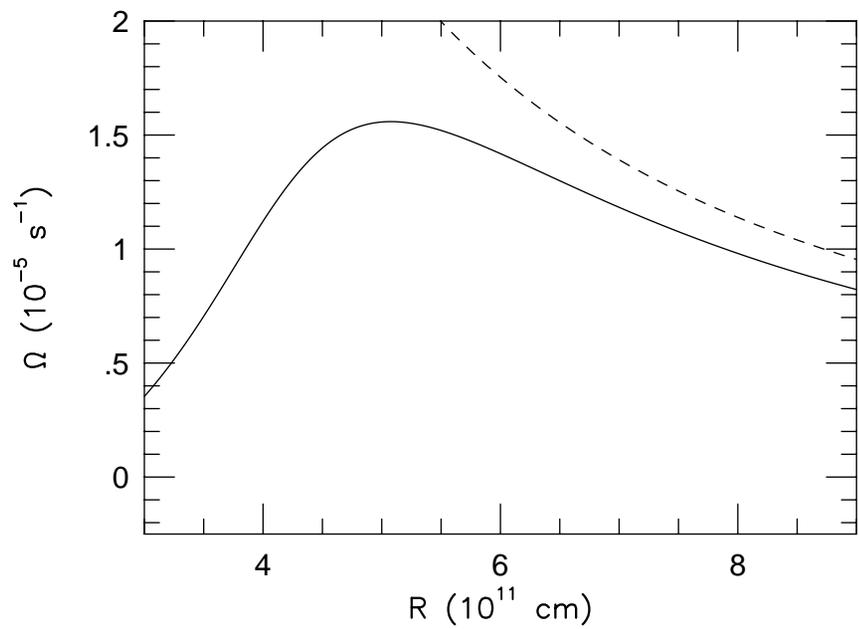
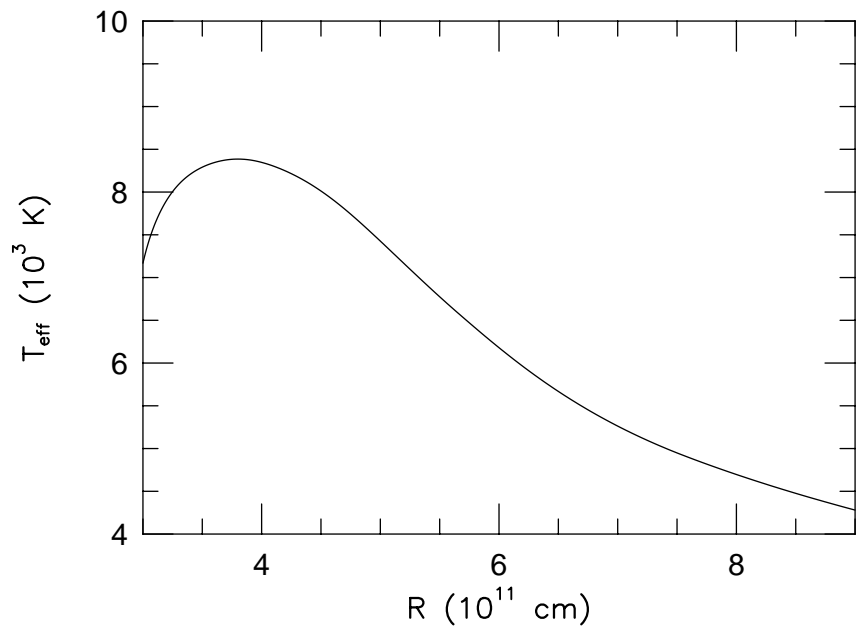
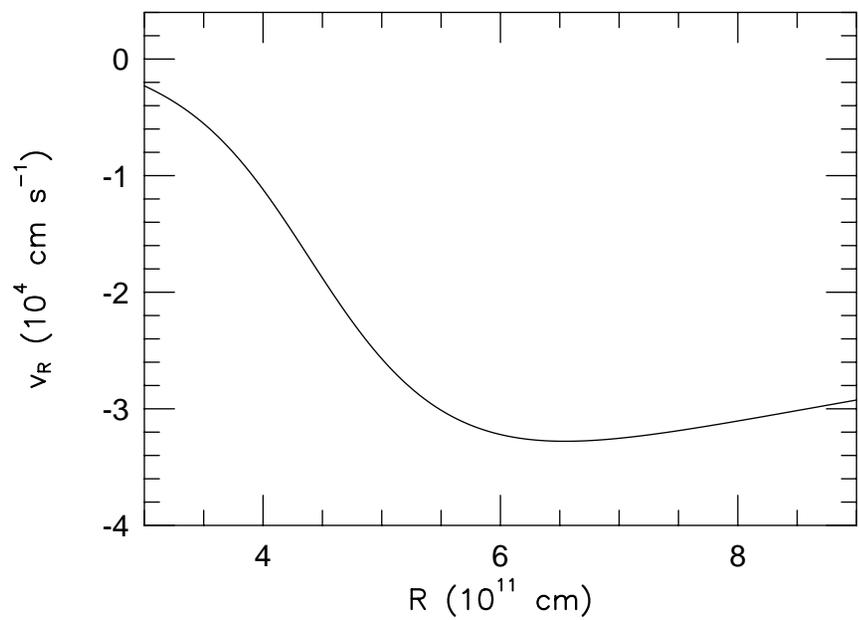
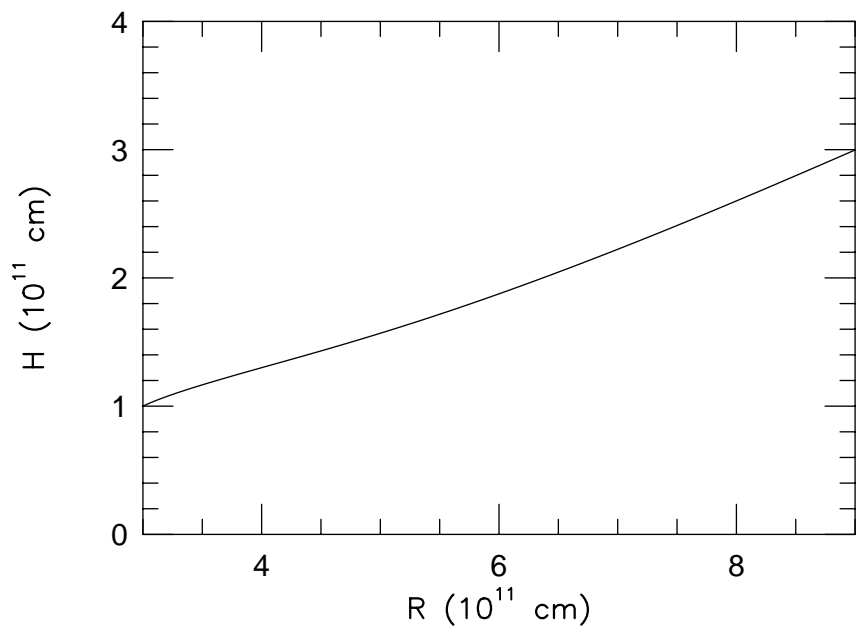
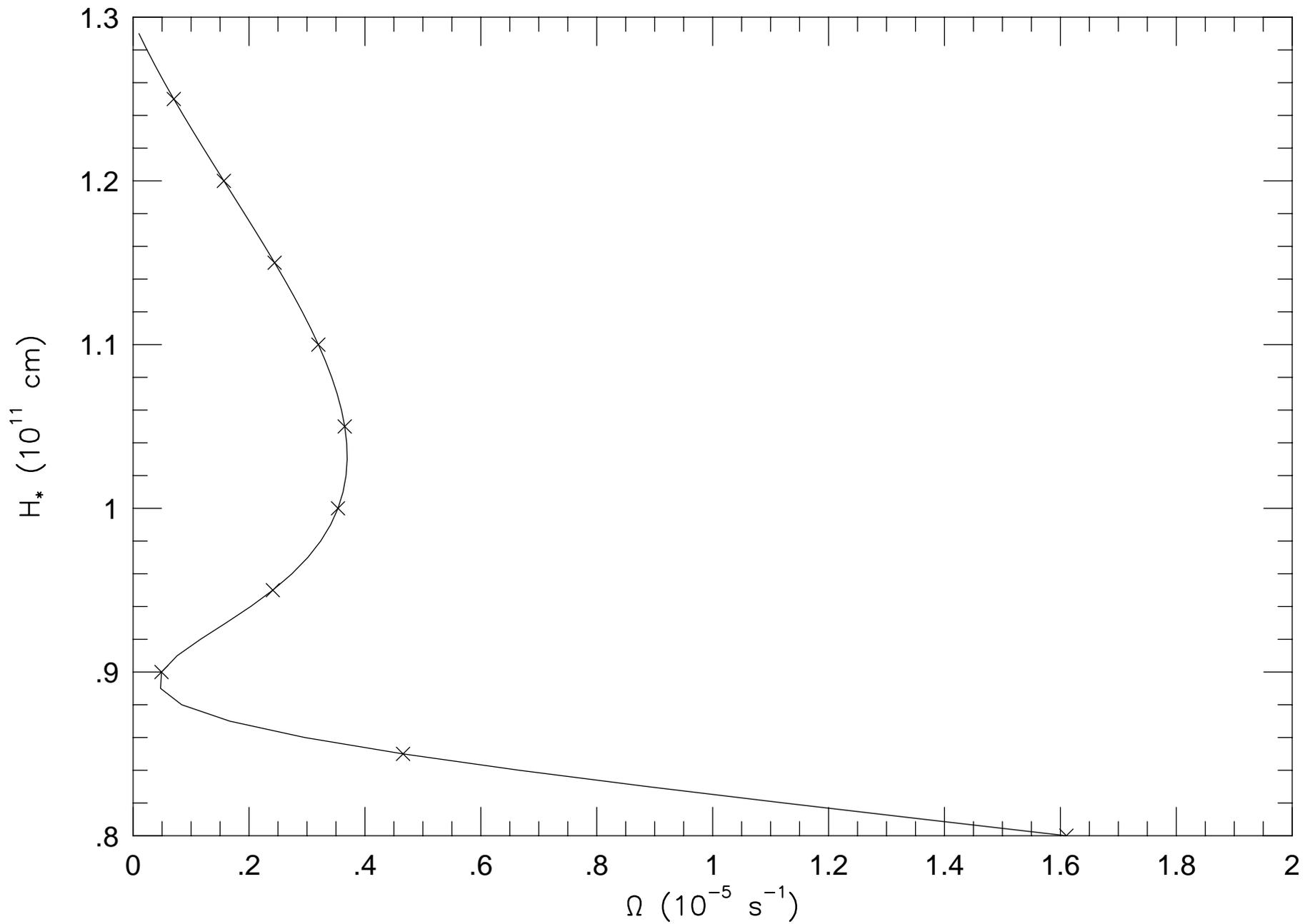
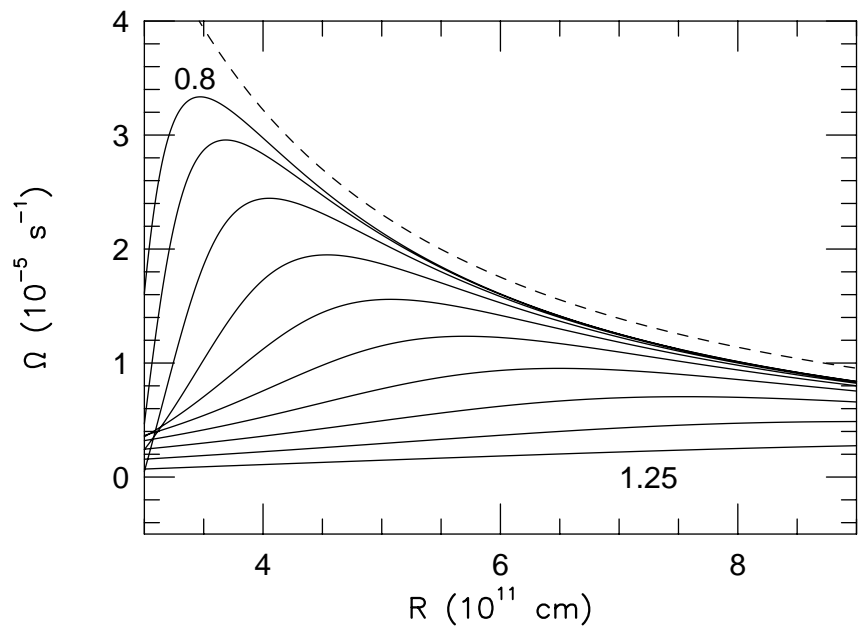
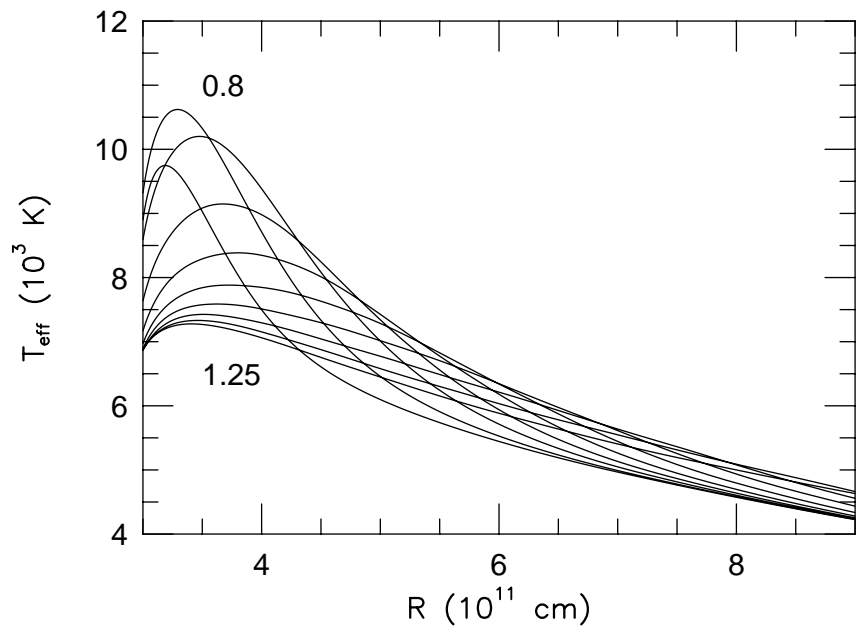
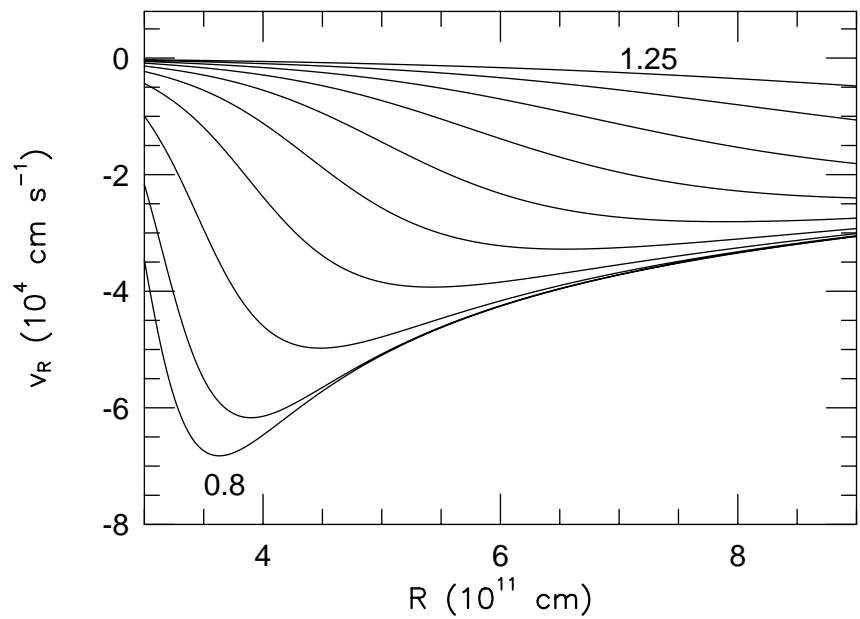
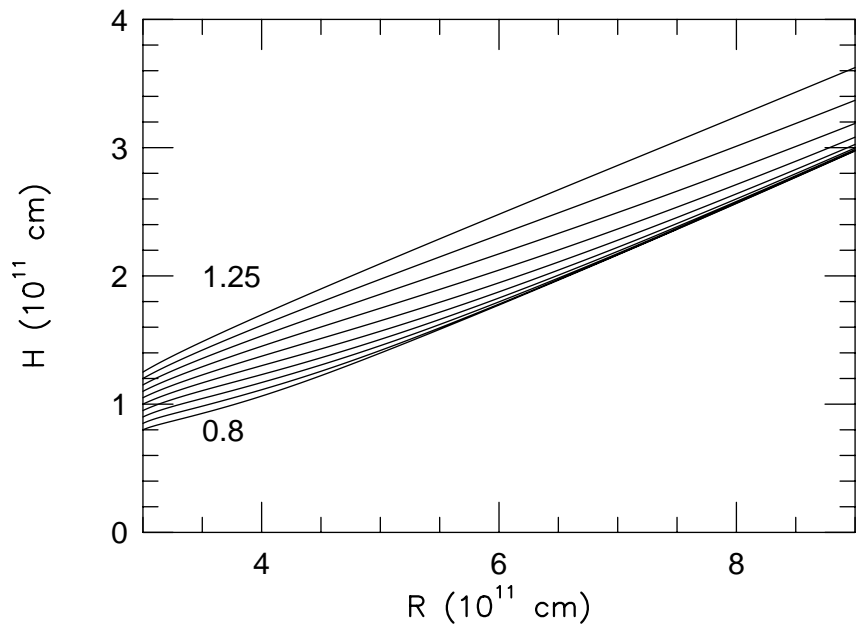
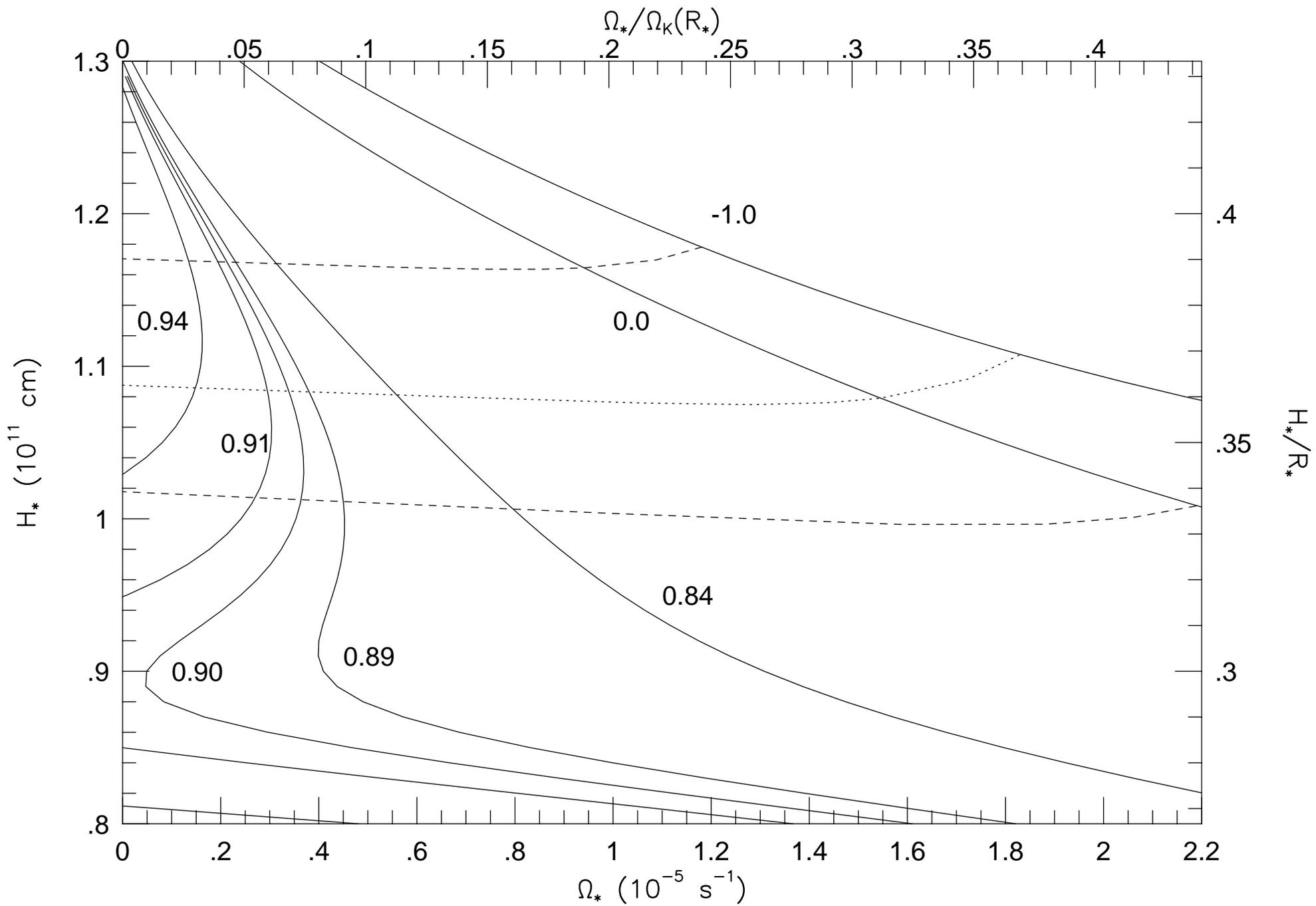
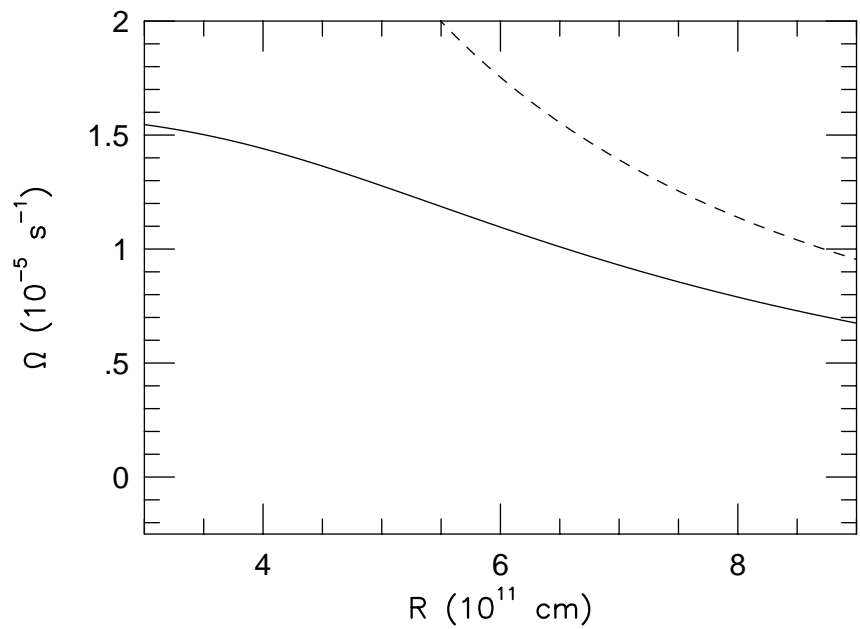
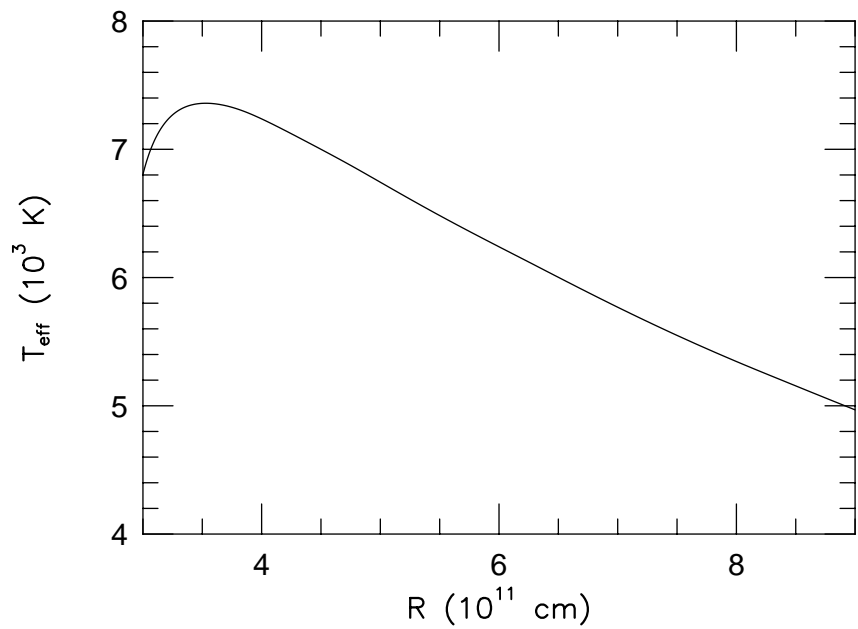
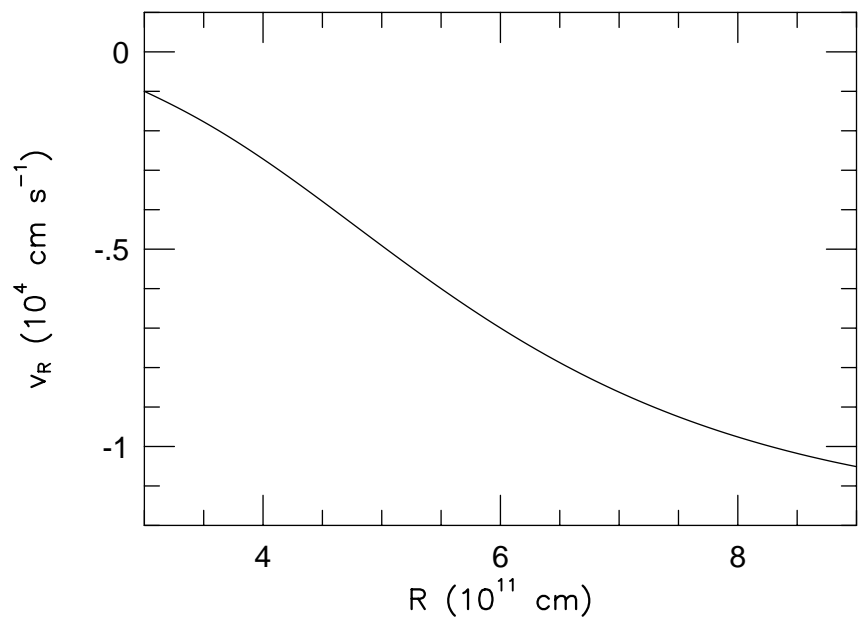
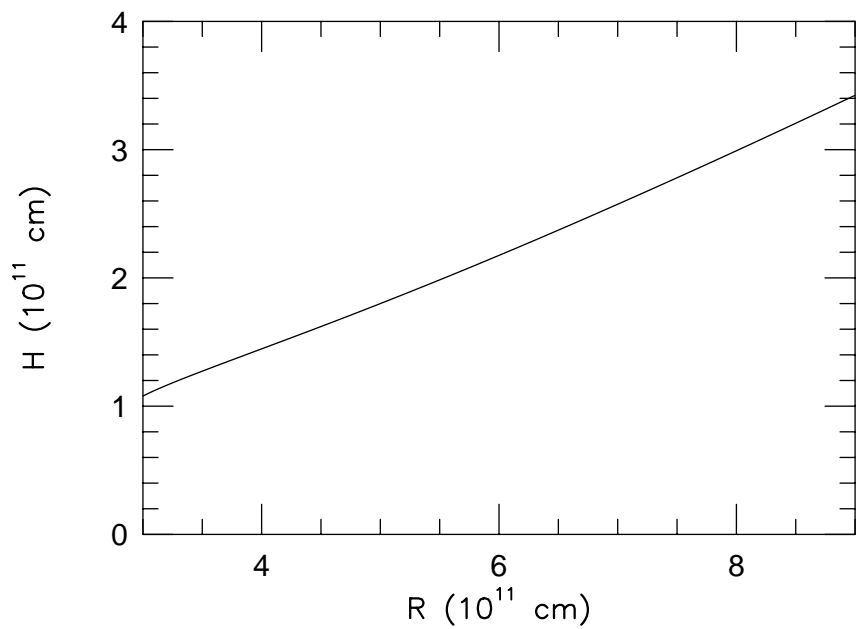
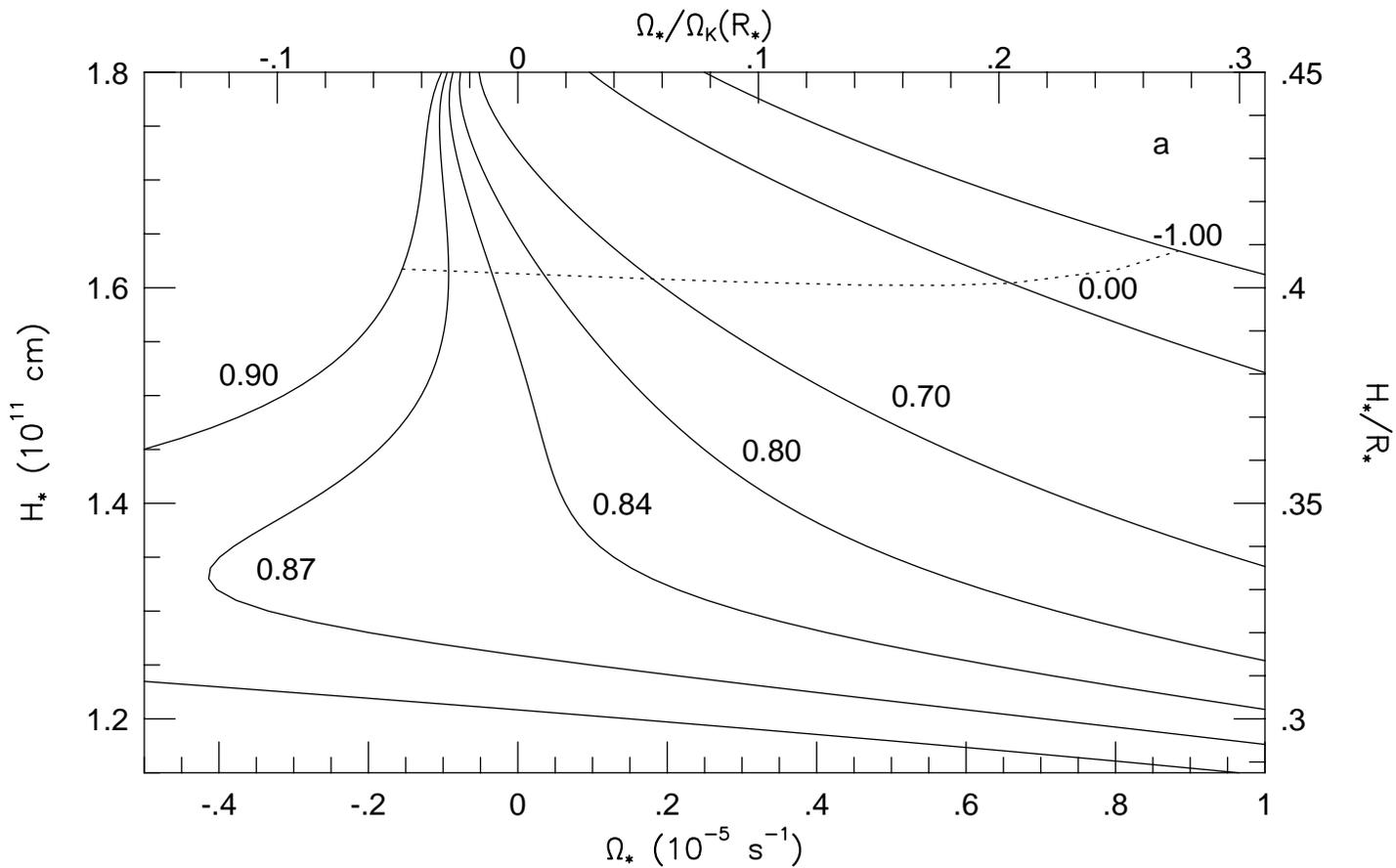
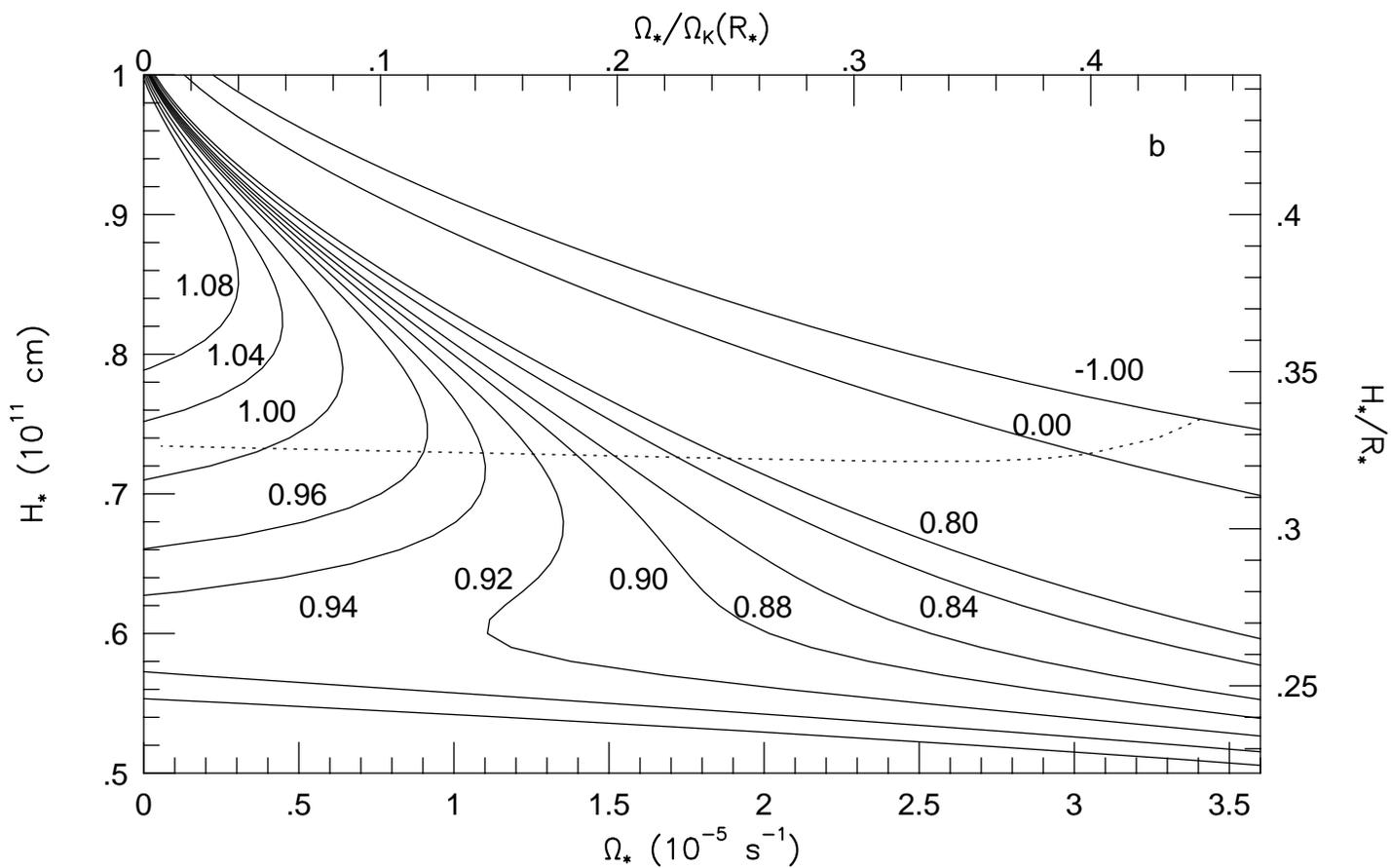
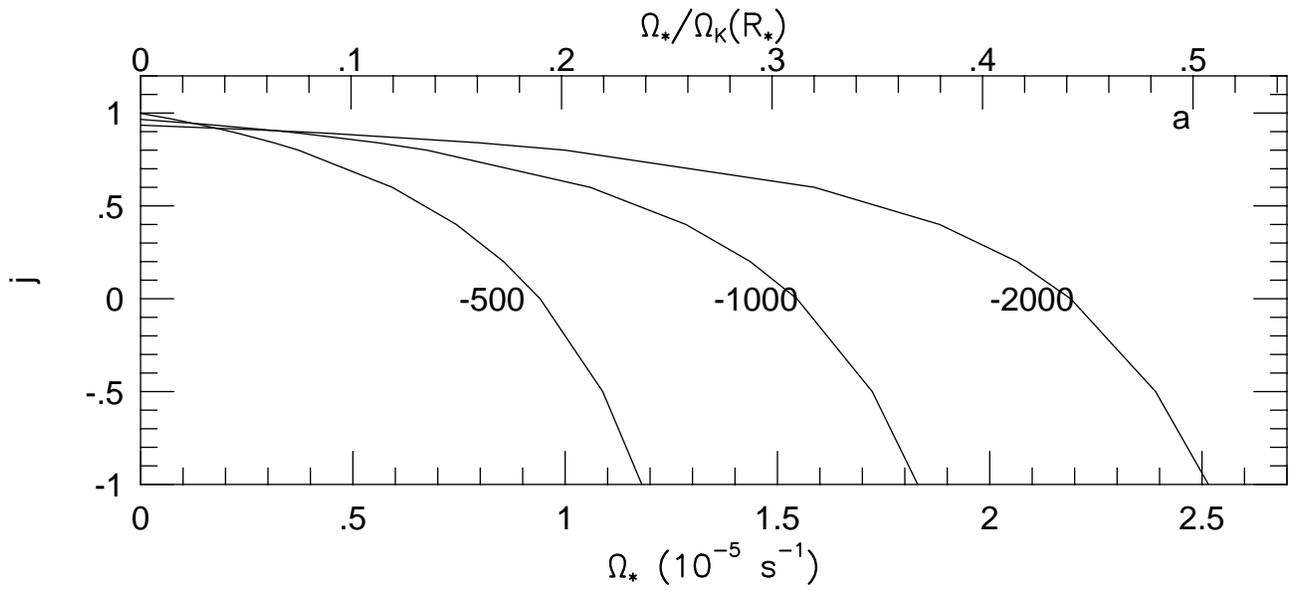
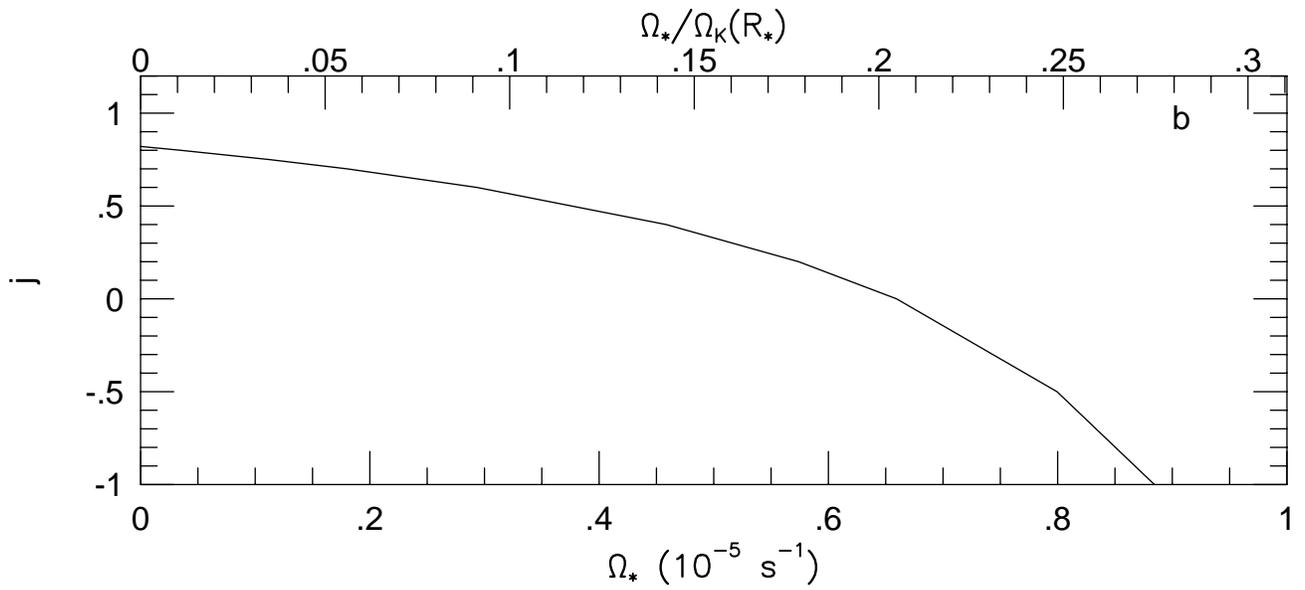
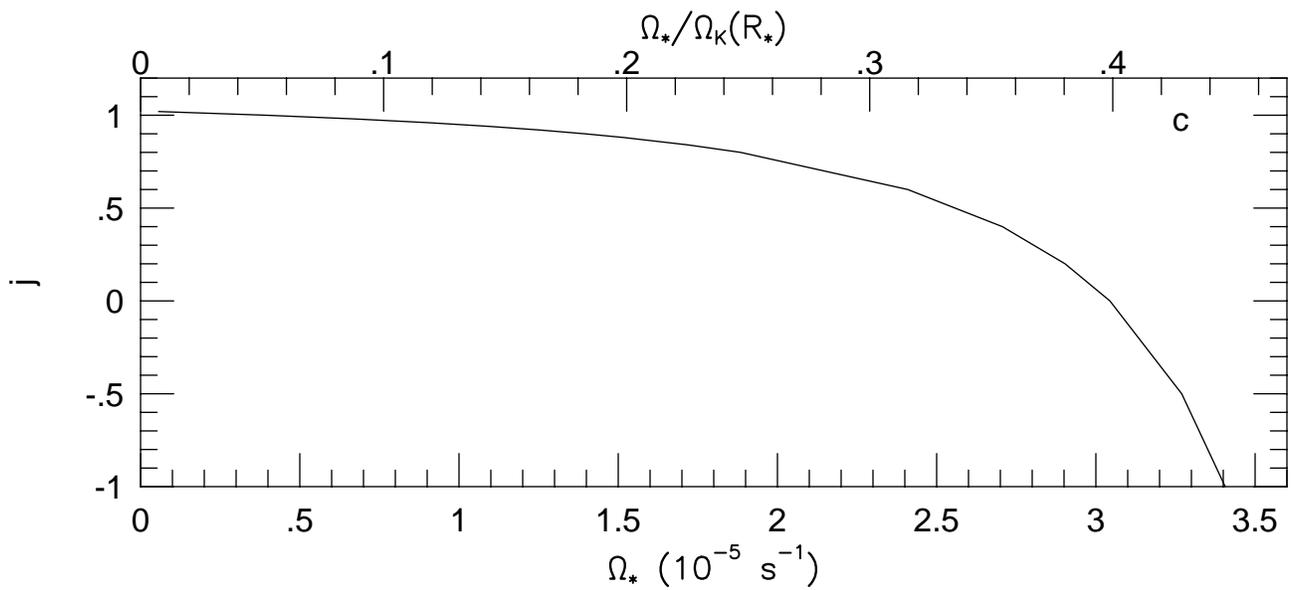
\end{document}